\documentclass[11pt,a4paper]{article}
\pdfoutput=1
\usepackage{jheppub}
\usepackage{color}
\usepackage{amsmath}

\def\s{{\mathbf s}}
\def\q{{\mathbf q}}

\def\k{{\mathbf k}}
\def\x{{\mathbf x}}
\def\y{{\mathbf y}}

\def\r{{\mathbf r}}
\def\z{{\mathbf z}}

\def\v{{\mathbf v}}
\def\u{{\mathbf u}}
\def\b{{\mathbf b}}
\def\B{{\mathbf B}}
\def\G{{\cal G}}
\def\U{\mathcal{U}}

\def\P{{\cal P}}

\def\ux{{\underline{x}}}
\def\uy{{\underline{y}}}
\def\uz{{\underline{z}}}
\def\uk{{\underline{k}}}
\def\R{{\cal R}}
\def\A{{\cal A}}
\def\del{{\partial}}
\def\s{{\mathbf s}}
\def\t{{\mathbf t}}
\def\hz{{\hat{\mathbf z}}}
\def\mz{{\mathfrak z}}
\def\C{{\cal C}}
\def\tC{{\tilde{\cal C}}}
\def\O{{\cal O}}

\newcommand{\beq}{\begin{eqnarray}}
\newcommand{\eeq}{\end{eqnarray}}
\newcommand{\be}{\begin{eqnarray*}}
\newcommand{\ee}{\end{eqnarray*}}

\DeclareMathOperator{\tr}{tr}

\title{Next-to-next-to-eikonal corrections in the CGC}
\author[a]{Tolga Altinoluk,}
\author[a]{N\'estor Armesto,}
\author[b]{Guillaume Beuf}
\author[a]{and Alexis Moscoso}
\affiliation[a]{Departamento de F\'isica de Part\'iculas and IGFAE,
Universidade de Santiago de Compostela,
E-15706 Santiago de Compostela,
Galicia-Spain}
\affiliation[b]{Department of Physics, Ben-Gurion University of the Negev,
Beer Sheva 84105, Israel}

\emailAdd{tolga.altinoluk@usc.es}
\emailAdd{nestor.armesto@usc.es}
\emailAdd{beuf@post.bgu.ac.il}
\emailAdd{alexis.mrial@gmail.com}

\abstract{
We extend the study of corrections to the eikonal approximation that was initiated in Ref. \cite{Altinoluk:2014oxa} to higher orders. These corrections associated with the finite width of the target are investigated 
and the gluon propagator in background field is calculated at next-to-next-to-eikonal accuracy. The result is then applied to the single inclusive gluon production cross section at central rapidities and the light-front helicity asymmetry, in pA collisions, in order to analyse these observables beyond the eikonal limit.
The next-to-next-to-eikonal corrections to the unpolarized cross section are non-zero and provide the first corrections to the usual $k_\perp$-factorized expression.
In contrast, the eikonal and next-to-next-to-eikonal contributions to the helicity asymmetry vanish, while the next-to-eikonal ones are non-zero.

}

\begin{document}
\maketitle

\pagestyle{empty}
\newpage

\mbox{}

\pagestyle{plain}

\setcounter{page}{1}

\section{Introduction}
\label{sec:intro}
%


Particle production in high-energy collisions involving nuclei constitutes a hot topic of research in Quantum Chromodynamics (QCD) due to both its theoretical and practical importance. Depending on the size of the relevant scale associated to the production process, extensions of collinear factorization (see \cite{Collins:1989gx} and references therein) for a truly hard scale, or of $k_\perp$-factorization \cite{Catani:1990eg,Levin:1991ry} for a semi-hard scale, are demanded.

The existing calculations for single inclusive particle production  in pA collisions - a situation usually referred to as dilute-dense scattering - result in a $k_\perp$-factorized expression \cite{Kovchegov:1998bi,Kovchegov:2001sc,Dumitru:2001ux,Kharzeev:2003wz,Blaizot:2004wu,Blaizot:2004wv} at the lowest order in the coupling constant if the particle is produced at central rapidity. By contrast, the hybrid factorization \cite{Dumitru:2005gt} is relevant if the particle is produced at forward rapidity\footnote{There exists a large activity on computing and understanding the corresponding next-to-leading corrections in the coupling constant \cite{Altinoluk:2011qy,Chirilli:2012jd,Stasto:2013cha,Stasto:2014sea,Kang:2014lha,Xiao:2014uba,Altinoluk:2014eka}.}. All these calculations rely on: (a) the replacement of the target by a large background field; and (b) the use of the eikonal approximation, in which the constituents of the projectile just experience color rotation upon scattering with the target through picking a Wilson line at a given transverse point but integrated along the light-cone
direction of propagation of the projectile, for which the target appears as infinitely Lorentz contracted. The  Color Glass Condensate (CGC, see \cite{Gelis:2010nm} and references therein) employs both assumptions for resumming high-energy logarithms, but neglects corrections suppressed by the energy of the collision, a situation parallel to neglecting the inverse powers of the hard scale in collinear factorization.

In the previous work \cite{Altinoluk:2014oxa} we developed a systematic method to compute higher order corrections to the eikonal approximation in the CGC (see related work in the context of soft gluon resummation for hard scattering amplitudes in \cite{Laenen:2008gt,Laenen:2010uz}). Such next-to-eikonal contributions, due to the non-zero length of the target or the finiteness of the energy of the projectile, can be understood as the subleading effects with respect to either the infinite Lorentz contraction of the target or the infinite Lorentz dilation of the projectile.
The method involves new operators  that contain transverse gradients of the background field in the target, which are thus sensitive to its transverse structure. We applied such formalism to single-inclusive particle production at central rapidities, recovering the usual $k_\perp$-factorized formula in pA collisions and obtaining the first next-to-eikonal correction. For collisions in which no privileged direction exist, the next-to-eikonal correction was found to vanish identically, thus extending the validity of the usual $k_\perp$-factorized formula. On the contrary, for collisions in which there is some privileged direction like in the study of spin asymmetries for a polarized target, it was the next-to-eikonal correction the first non-vanishing contribution. This resembles the situation in collinear factorisation where the first power-suppressed contributions become dominant for semi-inclusive production or polarized collisions, see e.g. \cite{Efremov:1984ip,Qiu:1991pp,Qiu:1998ia,Bacchetta:2006tn,Brodsky:2002cx,Barone:2010zz,Boer:2011fh} and references therein.

In the present paper we extend the formalism of \cite{Altinoluk:2014oxa} to next-to-next-to-eikonal accuracy.  In Section \ref{retarded} we compute the corresponding retarded gluon propagator that will be used subsequently to compute particle spectra. In Section \ref{spectra} we calculate both unpolarized and polarized single inclusive gluon production in unpolarized pA collisions. Finally we present our conclusions, that we anticipate here: the computed next-to-next-to-eikonal corrections for
the gluon helicity asymmetry
 vanish identically
, while those for unpolarized production are non-zero and provide the first corrections to the $k_\perp$-factorized expression  \cite{Kovchegov:1998bi,Kovchegov:2001sc,Dumitru:2001ux,Kharzeev:2003wz,Blaizot:2004wu,Blaizot:2004wv}. These results extend the conclusions extracted in the previous work \cite{Altinoluk:2014oxa} to higher orders. New operators appear, whose meaning is discussed and whose rapidity evolution is left for future studies\footnote{Note that the rapidity evolution of these decorated dipole operators may be related to the resummations recently discussed in the context of jet quenching in Refs. \cite{Liou:2013qya,Blaizot:2014bha,Iancu:2014kga,Abir:2015qva}.}.

\section{Retarded gluon propagator in background field at next-to-next-to-eikonal accuracy}
\label{retarded}

In the calculation of high-energy dilute-dense scattering processes, for example within the CGC formalism, the most important building blocks are the gluon and quark propagators in a strong classical background gluon field representing the dense target. In this paper we restrict ourselves to the gluon propagator. We choose a frame in which the target is a highly boosted left-mover, and pick the light-cone gauge, $A^+_a=0$.

In the limit of large Lorentz boost, one can in principle perform the following approximations to the classical background field $\A^{\mu}_a(x)$:
\begin{eqnarray}
\A^{\mu}_a(x) &\simeq & \delta^{\mu -}\; \A^{-}_a(x)\label{Eik_app_1},\\
\A^{\mu}_a(x) &\simeq & \A^{\mu}_a(x^+, \x)\label{Eik_app_2},\\
\A^{\mu}_a(x) &\propto & \delta(x^+)\label{Eik_app_3}\, .
\end{eqnarray}
These approximations are together equivalent, at the level of the background field, to the eikonal approximation \cite{Bjorken:1970ah,Collins:1985gm} usually formulated at the level of the scattering amplitude.
The second and third approximations correspond respectively to infinite time dilation and infinite length contraction. Corrections to each of the three approximations (\ref{Eik_app_1}-\ref{Eik_app_3}) are suppressed by inverse powers of the Lorentz boost of the target, which correspond to inverse powers of the total energy at the level of a high-energy scattering amplitude for a dilute projectile off that background field.

In the present paper, as in Ref. \cite{Altinoluk:2014oxa}, we investigate corrections to the eikonal approximation related only to the violation of the approximation \eqref{Eik_app_3}, i.e. associated with the finite length of the target. Although the three approximations (\ref{Eik_app_1}-\ref{Eik_app_3}) have a priori the same validity range, it can be justified to relax only the approximation \eqref{Eik_app_3} when the target is a large nucleus, since the $A^{1/3}$ enhancement of the length can compensate the suppression by the boost factor.

Hence, in the rest of the present section, we will study the retarded gluon propagator $G^{\mu\nu}_{R}(x,y)_{ab}$ in the presence of a classical gluon background field
\begin{equation}
\A^{\mu}_a(x)\equiv   \delta^{\mu -}\; \A^{-}_a(x^+, \x)\label{Background_field_def}\, ,
\end{equation}
with an arbitrary $x^+$ and $\x$ dependence.


\subsection{From gluon to scalar propagator in the background field}

In the light-cone gauge, the gluon propagator obeys the constraints
\begin{equation}
G^{+\nu}_{R}(x,y)_{ab}=G^{\mu+}_{R}(x,y)_{ab}=0\, .
\end{equation}
Due to the $x^-$ independence of the background field \eqref{Background_field_def}, it is useful to take the Fourier transform from $x^-$ to $k^+$, and define
\begin{equation}
G^{\mu\nu}_{R}(x,y)_{ab}=\int \frac{\textrm{d}p^+}{2\pi}\, e^{-ik^+ (x^-\!-y^-)}\; \frac{1}{2 (k^+\!+\!i\epsilon)}  \,{\cal G}^{\mu\nu}_{k^+}(\underline{x};\underline{y})_{ab}\, ,\label{FTPropMixed}
\end{equation}
using the notation $\underline{x}=(x^+,\x)$.

Linearizing the classical Yang-Mills equations around the background field $\A^{-}_a(x^+, \x)$, one can find the Green's equations obeyed by the background propagator $G^{\mu\nu}_{R}(x,y)_{ab}$. Using the Fourier representation \eqref{FTPropMixed}, the solution of these Green's equations can be given in terms of a retarded scalar propagator in background field, as
\begin{eqnarray}
{\cal G}^{ij}_{k^+}(\underline{x};\underline{y})^{ab}&=&\delta^{ij}\; {\cal G}^{ab}_{k^+}(\underline{x};\underline{y}),\label{VectPerpPerp2ScalMix}\\
{\cal G}^{-i}_{k^+}(\underline{x};\underline{y})^{ab}&=&\frac{-i}{k^+\!+\!i\epsilon}\;  \partial_{\x^i}\,  {\cal G}^{ab}_{k^+}(\underline{x};\underline{y}),\label{VectMinPerp2ScalMix}\\
{\cal G}^{i-}_{k^+}(\underline{x};\underline{y})^{ab}&=&\frac{i}{k^+\!+\!i\epsilon}\;  \partial_{\y^i}\,  {\cal G}^{ab}_{k^+}(\underline{x};\underline{y}),\label{VectPerpMin2ScalMix}\\
{\cal G}^{--}_{k^+}(\underline{x};\underline{y})^{ab}&=&\!\!\frac{1}{(k^+\!+\!i\epsilon)^2}\;  \partial_{\x^i}\,\partial_{\y^i}\,  {\cal G}^{ab}_{k^+}(\underline{x};\underline{y})\!+\!\frac{2i}{k^+\!+\!i\epsilon}\, \delta^{ab}\;  \delta^{(3)}\!(\underline{x}\!-\!\underline{y})  \label{VectMinMin2ScalMix}\, ,
\end{eqnarray}
where the scalar propagator ${\cal G}^{ab}_{k^+}(\underline{x};\underline{y})$ is the retarded solution of the Green's equation
\begin{equation}
\bigg[\delta^{ab} \,\left(i\partial_{x^+}+\frac{\partial_{\x}^2}{2(k^+\!+\!i\epsilon)}
\right)   +g \Big({\cal A}^{-}(\underline{x})\cdot T\Big)^{\! ab} \bigg] {\cal G}^{bc}_{k^+}(\underline{x};\underline{y})= i\, \delta^{ac}\: \delta^{(3)}(\underline{x}\!-\!\underline{y})\, . \label{SchroEq}
\end{equation}
We refer the reader to Ref. \cite{Altinoluk:2014oxa} for more details on the procedure just described.

The interpretation of the solution (\ref{VectPerpPerp2ScalMix}-\ref{VectMinMin2ScalMix}) for the gluon propagator in background field is the following. Due to the approximations \eqref{Eik_app_1} and \eqref{Eik_app_2}, interactions with the background field $\A^{-}_a(x^+, \x)$ cannot change the polarization state of a gluon. Hence, the two physical polarizations of the gluon propagate independently through the medium, each of them according to the scalar equation \eqref{SchroEq}. Then, the additional term with $\delta^{(3)}\!(\underline{x}\!-\!\underline{y})$ in eq. \eqref{VectMinMin2ScalMix} accounts for the non-propagating polarization associated with purely Coulombian interaction.

The scalar Green's equation \eqref{SchroEq} has a form of a Schr\"odinger equation in $2+1$ dimensions with a matrix potential. Its solution can thus be written as a path integral~\cite{Baier:1996kr,Zakharov:1996fv,Zakharov:1998sv,Wiedemann:2000za,MehtarTani:2006xq}. For our purposes, we need the discretized form of the path integral, which reads
\beq
\G^{ab}_{k^+}(\ux,\uy) &=&  \lim_{N\to\infty} \int \left( \prod_{n=1}^{N-1}d^2\z_{n}\right) \left[ \prod_{n=0}^{N-1}\G_{0,k^+}(z^+_{n+1},\z_{n+1};z^+_n,\z_n)\right] \U^{ab} \left( x^+,y^+; \{\z_n\}\right)\, ,\nonumber \\
&& \label{disc_scalar_prop_def}
\eeq
with the boundary conditions $\z_{0}\equiv \y$ and $\z_{N}\equiv \x$, and
\beq
z^+_n=y^++\frac{n}{N}(x^+-y^+) \label{z_n_plus}\, .
\eeq
The free scalar propagator appearing in the path integral expression, given in eq. \eqref{disc_scalar_prop_def}, reads
\beq
\G_{0,k^+}(\ux,\uy) &=& \theta(x^+\!-\!y^+)\; \left(\frac{-ik^+}{2\pi(x^+\!-\!y^+)}\right)\;
\exp\left( \frac{ik^+}{2(x^+\!-\!y^+)}(\x-\y)^2\right)\, .\label{Free_scalar_propag}
\eeq
The discretized Wilson line is defined as
\begin{eqnarray}
\U^{ab}(x^+,y^+,\{\z_n\})&=& {\cal P}_{+} \Bigg\{\prod_{n=0}^{N-1}  \exp \bigg[ig \frac{(x^+\!-\!y^+)}{N}   \Big({\cal A}^{-}(z^+_{n},\z_{n})\cdot T\Big)\bigg]\Bigg\}^{ab}
\, ,  \label{DiscrWilsonLine}
\end{eqnarray}
with ${\cal P}_{+}$ denoting path ordering along the $x^+$ direction.

In the expression of observables related to dilute-dense scattering at high energy, the background propagator appears typically through its transverse Fourier transform $\int d^2\x \; e^{-i\k\cdot\x} \; \G^{ab}_{k^+}(\ux,\uy)$.
In order to study corrections with respect to the eikonal approximation (or more precisely, with respect to the shockwave approximation \eqref{Eik_app_3}) for the background gluon propagator, we start from that Fourier representation of the background scalar propagator, and define the corresponding medium modification factor $\widetilde{\R}^{ab}_{\uk}(x^+,y^+;\y)$
as
\beq
\int d^2\x \; e^{-i\k\cdot\x} \; \G^{ab}_{k^+}(\ux,\uy)&=&  \widetilde{\R}^{ab}_{\uk}(x^+,y^+;\y)\; \int d^2\x \; e^{-i\k\cdot\x} \; \G_{0,k^+}(\ux,\uy)\nonumber\\
&=& \widetilde{\R}^{ab}_{\uk}(x^+,y^+;\y)\;\; \theta(x^+\!-\!y^+)\; e^{-i\k\cdot\y}\; e^{\frac{-i(x^+-y^+)\k^2}{2k^+}}\; .\label{Rtilde_def}
\eeq

In order to calculate the expansion of $\widetilde{\R}^{ab}_{\uk}(x^+,y^+;\y)$ beyond the eikonal approximation, it is more convenient in practice to consider the large $k^+$ limit than the limit of large boost of the target. The two limits are of course equivalent, thanks to the invariance of the whole scattering process under longitudinal boosts.

There are two ways to define the large $k^+$ limit: either it can be taken at fixed $\k$, or at fixed $\k/k^+$. The former case is relevant for high-energy scattering processes in the Regge limit, whereas the latter is relevant for hard scattering processes in the Bjorken limit. The variable $\k/k^+$ is related to the deflection angle for the gluon. One considers small angle scattering in the first case and finite angle scattering in the second, in the large $k^+$ limit.

Both for convenience, and in order to obtain results applicable in each case, we perform the large $k^+$ expansion of $\widetilde{\R}^{ab}_{\uk}(x^+,y^+;\y)$ in two steps. First, we calculate the $k^+$ expansion at fixed $\k/k^+$ in the section \ref{sec:semi_classical_exp}, and then we re-expand the result for small $\k/k^+$ in the section \ref{sec:small_angle_exp}.


\subsection{Semi-classical expansion for the background propagator \label{sec:semi_classical_exp}}

In the large $k^+$ limit at fixed $\k/k^+$, the path integral for the gluon propagator becomes increasingly dominated by the classical trajectory
\beq
\label{zhat_cont}
\hat{\z}(z^+)=\y+(z^+-y^+)\frac{\k}{k^+}\, .
\eeq
Our method to calculate the expansion of $\widetilde{\R}^{ab}_{\uk}(x^+,y^+;\y)$ in this limit is the following.
First, we take the Fourier transform $\x\mapsto \k$ of the discretized expression \eqref{disc_scalar_prop_def} for the propagator and change variables in order to write the path integral as an integral over deviations with respect to the classical path \eqref{zhat_cont}, making evident the factorization of the free propagator contribution as in eq. \eqref{Rtilde_def}. Second, we take the continuum limit. Third, we Taylor-expand insertions of the background field around the classical path, and perform explicitly the path integration order by order, in order to get the eikonal expansion at finite angle.


\subsubsection{Discretized form of the scalar medium modification factor}

Taking the Fourier transform $\x\mapsto \k$ of eq. \eqref{disc_scalar_prop_def} and relabeling $\x\to\z_N$, one gets
\beq
\label{disc_scalar_prop_FT}
\int d^2\x \; e^{-i\k\cdot\x} \; \G^{ab}_{k^+}(\ux,\uy)& = &\lim_{N\to\infty}  \int \left( \prod_{n=1}^{N}d^2\z_{n}\right)\left(\frac{-ik^+N}{2\pi(x^+-y^+)}\right)^N \\
&&\hspace{-2.5cm}\times \;e^{-i\k\cdot\z_N}\left[ \prod_{n=0}^{N-1}\exp\left( \frac{ik^+N}{2(x^+-y^+)}(\z_{n+1}-\z_n)^2\right)\right] \U^{ab} \left( x^+,y^+; \{\z_n\}\right)\, , \nonumber
\eeq
where $\z_0\equiv \y$.
The next step is to perform the change of variables $\z_n\mapsto \u_n = \z_n- \hat{\z}_n$, where $\hat{\z}_n$ is the discretized version of the classical trajectory, i.e.
\beq
\label{zhat_discr}
\hat{\z}_n=\y+(z^+_n-y^+)\frac{\k}{k^+}\, .
\eeq
Noting that
\beq
\hspace{-1cm}(\z_{n+1}-\z_n)^2&=&\left[(z^+_{n+1}-z^+_n)\frac{\k}{k^+}+\u_{n+1}-\u_n\right]^2=\left[\frac{1}{N}(x^+-y^+)\frac{\k}{k^+}+\u_{n+1}-\u_n\right]^2\\
&=&(\u_{n+1}-\u_n)^2+\frac{2}{N}(x^+-y^+)\frac{\k}{k^+}\cdot(\u_{n+1}-\u_n)+\frac{(x^+-y^+)^2}{N^2}\left(\frac{\k}{k^+}\right)^2, \nonumber
\eeq
one finds that
\beq
 e^{-i\k\cdot\z_N}\left[ \prod_{n=0}^{N-1}\exp\left( \frac{ik^+N}{2(x^+-y^+)}(\z_{n+1}-\z_n)^2\right)\right]
 &=& e^{-i\k\cdot\y-i \frac{(x^+-y^+)\k^2}{k^+}-i\k\cdot\u_N} \nonumber\\
& &\hspace{-6cm} \times
 \left[ \prod_{n=0}^{N-1}\exp\left( \frac{ik^+N}{2(x^+-y^+)}(\u_{n+1}-\u_n)^2\right)\right]
 \;  e^{i\k\cdot(\u_N-\u_0)+i \frac{(x^+-y^+)\k^2}{2k^+}}\nonumber\\
& & \hspace{-6cm} = e^{-i\k\cdot\y}\;  e^{-i \frac{(x^+-y^+)\k^2}{2k^+}} \;  \left[ \prod_{n=0}^{N-1}\exp\left( \frac{ik^+N}{2(x^+-y^+)}(\u_{n+1}-\u_n)^2\right)\right]\, ,\label{phases_path_int}
\eeq
because $\u_0\equiv 0$. From the equations \eqref{Rtilde_def}, \eqref{disc_scalar_prop_FT} and \eqref{phases_path_int}, one obtains the discretized path-integral expression for the medium-modification factor $\widetilde{\R}^{ab}_{\uk}(x^+,y^+;\y)$ as
\beq
\label{Rtilde_path_int}
\widetilde{\R}^{ab}_{\uk}(x^+,y^+;\y)&=&\lim_{N\to\infty}\int\left(\prod_{n=1}^{N}d^2\u_n \right) \Bigg\{  \prod^{N-1}_{n=0} \G_{0,k^+}(z^+_{n+1},\u_{n+1};z^+_n,\u_n)\Bigg\} \nonumber\\
&&\hspace{3cm} \times \U^{ab}\left(x^+,y^+;\left\{\hat{\z}_n+\u_n\right\}\right) \, .
\eeq


\subsubsection{Continuum limit}

Let us remind the Lie product formula which is valid for any square matrices $A$ and $B$:
\beq
e^{A+B}=\lim_{l\to+\infty}\left(e^{\frac{A}{l}}\cdot e^{\frac{B}{l}}\right)^l\,.
\eeq
For small $B$, $e^{\frac{B}{l}}=1+O(B)$. Thus, for small $B$, the Lie product formula can be written as
\beq
e^{A+B}&=&\lim_{l\to+\infty}\Bigg\{ e^A+ \sum_{j=1}^{l} e^{\frac{jA}{l}} \left(e^{\frac{B}{l}}-1\right) e^{\frac{(l-j)A}{l}} \\
&&\hspace{2cm} +\sum_{j=1}^{l-1} \sum_{i=1}^{l-j} e^{\frac{jA}{l}} \left(e^{\frac{B}{l}}-1\right) e^{\frac{iA}{l}} \left(e^{\frac{B}{l}}-1\right) e^{\frac{(l-j-i)A}{l}} +O(B^3)\Bigg\}\, . \nonumber
\eeq
In the large $l$ limit, each sum can be written as an integral and one obtains
\beq
\label{Lie_Formula_for_Large_l}
e^{A+B}&=&e^A+\int^1_0 ds\; e^{sA} \; B \; e^{(1-s)A}+\int_0^1ds\int_0^{1-s}du \; e^{sA} \;  B \; e^{uA} \; B \; e^{(1-s-u)A}\; +\; O(B^3) .\nonumber \\ &&
\eeq
Introducing the notations
\beq
a_n&\equiv&(x^+\!-\!y^+)\: ig T\cdot \A^-(z^+_n;\hat{\z}_n)\label{def_an}\\
b_n(\u_n)&\equiv& (x^+\!-\!y^+)\: ig T\cdot \left[\A^-(z^+_n;\hat{\z}_n+\u_n)-\A^-(z^+_n;\hat{\z}_n)\right]\, ,\label{def_bn}
\eeq
and using  eq. \eqref{Lie_Formula_for_Large_l},
one can express each elementary gauge link (see eq. \eqref{DiscrWilsonLine}) of the discretized Wilson line appearing in eq. \eqref{Rtilde_path_int} as
\beq
&& 
\exp\left[\frac{(x^+\!-\!y^+)}{N}\; ig\;T\cdot\A^-(z^+_n; \hat{\z}_n+\u_n)\right]
= e^{\frac{a_n + b_n(\u_n)}{N}}
\nonumber\\
&&
= e^{\frac{a_n}{N}}
+\int^1_0 ds\; e^{s \frac{a_n}{N}} \; \frac{b_n(\u_n)}{N} \; e^{(1-s) \frac{a_n}{N}}
\\
&&
+\int_0^1 ds \int_0^{1-s} du \; e^{s \frac{a_n}{N}} \;  \frac{b_n(\u_n)}{N} \; e^{u \frac{a_n}{N}} \; \frac{b_n(\u_n)}{N} \; e^{(1-s-u)\frac{a_n}{N}}\;
+\; O\left(\left(\frac{b_n(\u_n)}{N}\right)^3\right).\nonumber
\label{applied_Lie}
\eeq
This expansion can be inserted at each discrete step in eq. \eqref{Rtilde_path_int}.

In order to understand what happens in the continuum limit $N\rightarrow +\infty$, let us first consider the case in which only the leading term $e^{\frac{a_n}{N}}$ is kept at each step in eq. \eqref{Rtilde_path_int}, which we call the contribution $0$ to $\widetilde{\R}_{\uk}(x^+,y^+;\y)$. One has
\beq
\widetilde{\R}_{\uk}(x^+,y^+;\y)\bigg|_{0}&=& \lim_{N\to\infty} \int\left(\prod_{n=1}^{N}d^2\u_n \right) {\cal P}_{+}  \prod^{N-1}_{n=0} \Bigg\{ \G_{0,k^+}(z^+_{n+1},\u_{n+1};z^+_n,\u_n) \;\;
 e^{\frac{a_n}{N}}
 \Bigg\}
\nonumber\\
&=& \lim_{N\to\infty}  \U(x^+,y^+,\{\z_n\})
\nonumber\\
&=& \U\left( x^+,y^+; [\hat{\z}(z^+)]\right)
 \, ,
\eeq
using the identity
\beq
\label{n0}
\int d^2\u  \;\G_{0,k^+}(z^+,\u;v^+,\v) &=& \theta(z^+\!-\!v^+) \, ,
\eeq
and defining the Wilson line following the continuous trajectory $\hat{\z}(z^+)$ from $y^+$ to $x^+$ as
\begin{eqnarray}
\U(x^+,y^+,[\hat{\z}(z^+)])&\equiv & {\cal P}_{+}  \exp \bigg[ig  \int_{y^+}^{x^+}\!\!\!\!
dz^+\: T\cdot {\cal A}^{-}(z^+,\hat{\z}(z^+))\bigg]
\, .  \label{ContWilsonLine}
\end{eqnarray}

Next, let us consider the contribution $1$ to $\widetilde{\R}_{\uk}(x^+,y^+;\y)$, in which in one discrete step only the terms subleading with respect to $e^{\frac{a_n}{N}}$ in eq. \eqref{applied_Lie} are kept, whereas in all the other steps, only $e^{\frac{a_n}{N}}$ is kept. This contribution reads
\beq
& &\widetilde{\R}_{\uk}(x^+,y^+;\y)\bigg|_{1}= \lim_{N\to\infty} \int\left(\prod_{n=1}^{N}d^2\u_n \right) \Bigg\{  \prod^{N\!-\!1}_{n=0} \G_{0,k^+}(z^+_{n+1},\u_{n+1};z^+_n,\u_n)  \Bigg\}
\nonumber\\
&& 
\times \sum_{p=1}^{N\!-\!1} {\cal P}_{+}
 \Bigg\{\prod_{n=p+1}^{N\!-\!1}  e^{\frac{a_n}{N}} \Bigg\}\;
\Bigg\{\int^1_0 ds\; e^{s \frac{a_p}{N}} \; \frac{b_p(\u_p)}{N} \; e^{(1-s) \frac{a_p}{N}}
\nonumber\\
&& 
+\int_0^1 ds \int_0^{1-s} du \; e^{s \frac{a_p}{N}} \;  \frac{b_p(\u_p)}{N} \;
e^{u \frac{a_p}{N}} \; \frac{b_p(\u_p)}{N} \; e^{(1-s-u)\frac{a_p}{N}}\;
+\; O\left(\left(\frac{b_p(\u_p)}{N}\right)^3\right)
 \Bigg\}
\Bigg\{\prod_{n=0}^{p-1}  e^{\frac{a_n}{N}} \Bigg\}
\nonumber\\
&& = \lim_{N\to\infty}
\sum_{p=1}^{N\!-\!1} \int d^2\u_p\;  \G_{0,k^+}(z^+_{p},\u_{p};y^+,\u_{0})\;
{\cal P}_{+}
\Bigg\{\prod_{n=0}^{N\!-\!1}  e^{\frac{a_n}{N}} \Bigg\}\;
\Bigg\{\int^1_0 ds\; e^{s \frac{a_p}{N}} \; \frac{b_p(\u_p)}{N} \; e^{-s \frac{a_p}{N}}
\nonumber\\
&&
+\int_0^1 ds \int_0^{1-s} du \; e^{s \frac{a_p}{N}} \;  \frac{b_p(\u_p)}{N} \;
e^{u \frac{a_p}{N}} \; \frac{b_p(\u_p)}{N} \; e^{-(s+u)\frac{a_p}{N}}\;
+\; O\left(\left(\frac{b_p(\u_p)}{N}\right)^3\right)
\Bigg\}
\label{contrib_1}
\, ,
\eeq
using the identity given in eq. \eqref{n0} as well as the relation
\beq
\label{convolution_identity}
\int d^2\z\;   \G_{0,k^+}(\ux; \uz) \;\G_{0,k^+}(\uz;\uy)
&=& \theta(x^+\!-\!z^+)\; \theta(z^+\!-\!y^+)\;
\G_{0,k^+}(\ux;\uy)
 \, .
\eeq
In the continuum limit $N\rightarrow +\infty$, the sum over $p$ becomes an integration over $z^+$, with the correspondence
\beq
\frac{1}{N}\; \sum_{p=1}^{N\!-\!1}\quad \mapsto \quad \frac{1}{x^+\!-\!y^+}\; \int_{y^+}^{x^+}\!\!\!\! dz^+\, .
\eeq
Hence, the term linear in $b_p(\u_p)/N$ in eq. \eqref{contrib_1} gives a finite contribution in the continuum limit, whereas all the other ones converge to zero.
Moreover, one has $e^{s \frac{a_p}{N}}\rightarrow 1$, and thus the integration over $s$ becomes trivial. All in all, one gets
\beq
& &\widetilde{\R}_{\uk}(x^+,y^+;\y)\bigg|_{1} = \lim_{N\to\infty}
 \frac{1}{N}  \sum_{p=1}^{N\!-\!1} \int d^2\u_p\;  \G_{0,k^+}(z^+_{p},\u_{p};y^+,\u_{0})\; {\cal P}_{+} \Bigg\{b_p(\u_p)\; \prod_{n=0}^{N\!-\!1}  e^{\frac{a_n}{N}} \Bigg\}
\nonumber\\
&& = \int_{y^+}^{x^+}\!\!\!\! dz_1^+ \int d^2\u_1\;  \G_{0,k^+}(z^+_{1},\u_{1};y^+,\u_{0})
{\cal P}_{+} \Bigg\{ \U(x^+,y^+,[\hat{\z}(z^+)])\;
ig T\cdot \delta\A^-_1(\u_1) \Bigg\}
\label{contrib_1_final}
\, ,
\eeq
with the notation
\beq
\delta\A^-_n(\u_n)&\equiv&\A^-(z^+_n;\hat{\z}_n+\u_n)-\A^-(z^+_n;\hat{\z}_n)\, .\label{def_deltaA}
\eeq
Here, the $\P_+$ means that the decoration $ig T\cdot\delta\A^-_1(\u_1)$ has to be inserted into the Wilson line $\U\left(x^+,y^+; [\hat{\z}(z^+)]\right)$ at the appropriate place in order to fulfill the $x^+$ ordering.
That notation will be used frequently in the rest of the paper.

These results generalise easily to the leftover contributions to the medium modification factor $\widetilde{\R}_{\uk}(x^+,y^+;\y)$, with only the terms subleading with respect to $e^{\frac{a_n}{N}}$ in eq. \eqref{applied_Lie} kept in more than one discrete step, and only $e^{\frac{a_n}{N}}$ in the others. Adding all the contributions together, one obtains in the continuum limit
\beq
\label{finalRtilde}
\widetilde{\R}^{ab}_{\uk}(x^+,y^+;\y) &=& \U^{ab}\left(x^+,y^+;[\hat{\z}(z^+)]\right) +\sum_{l=1}^{+\infty} \int_{y^+}^{x^+} dz^+_1 \int_{z^+_1}^{x^+} dz^+_2 \cdots\int_{z^+_{l-1}}^{x^+}dz^+_l\int\left(\prod_{j=1}^{l}d^2\u_j\right)\nonumber\\
&&\hspace{-2cm}\times\left[\prod_{j=1}^{l}\G_{0,k^+}(z^+_j,\u_j;z^+_{j-1},\u_{j-1})\right]\P_+\Bigg\{\U\left(x^+,y^+;[\hat{\z}(z^+)]\right)\prod_{j=1}^ligT\cdot\delta\A^-_j(\u_j)\Bigg\}\, .
\eeq

\subsubsection{Expanding around the classical trajectory\label{sec:semi_class_exp}}

With the expression \eqref{finalRtilde}, we are now in the position to perform the semi-classical expansion of $\widetilde{\R}^{ab}_{\uk}(x^+,y^+;\y)$ around the straight trajectory $\hat{\z}(z^+)$ with a fixed angle associated with ${\k}/{k^+}$. The method is the following. First, one rewrites the finite difference $\delta\A^-_j(\u_j)$, defined in eq. \eqref{def_deltaA}, as a Taylor series in $\u_j$ every time it appears in eq. \eqref{finalRtilde}. Second, one can perform the integration over each of the transverse displacements $\u_j$. Upon integration, each power of $\u_j$ converts into a power of $1/\sqrt{k^+}$. This method can be pushed to arbitrary orders in principle, but becomes more
and more cumbersome at higher orders. In the present paper, we limit ourselves to the order $1/{k^+}^2$, corresponding to next-to-next-to-eikonal accuracy.
The details of the calculations are given in  appendix \ref{sec:app_semi_class_exp}.

For simplicity, let us introduce the following notations:
\beq
{\cal B}^i(\mz)&\equiv&igT\cdot \partial_{\y^i}\A^-(z^+, \hat{\z}(z^+)),\label{Bi}\\
{\cal B}^{ij}(\mz) &\equiv& igT\cdot \partial_{\y^i}\partial_{\y^j}\A^-(z^+, \hat{\z}(z^+)),\label{Bij}\\
{\cal B}^{ijl}(\mz) &\equiv& igT\cdot \partial_{\y^i}\partial_{\y^j}\partial_{\y^l}\A^-(z^+,\hz(z^+)),\label{Bijl}\\
{\cal B}^{ijlm}(\mz) &\equiv& igT\cdot \partial_{\y^i}\partial_{\y^j}\partial_{\y^l}\partial_{\y^m}\A^-(z^+, \hat{\z}(z^+))\, ,\label{Bijlm}
\eeq
where $\mz\equiv (z^+, \hat{\z}(z^+))$.

Collecting the results from Eqs. \eqref{Rl1final}, \eqref{Rl2final}, \eqref{Rl3final} and \eqref{Rl4final} in appendix \ref{sec:app_semi_class_exp}, one finds that the semiclassical expansion of the medium modification factor $\widetilde{\R}^{ab}_{\uk}(x^+,y^+;\y)$ reads
\beq
\label{R_semiclass_exp}
\hspace{-1cm}
\widetilde{\R}_{\uk}(x^+,y^+;\y)&=&\U(x^+,y^+;[\hat{\z}(z^+)]) +i\frac{(x^+\!-\!y^+)}{2k^+}\;\U_{\{1\}}(x^+,y^+;[\hat{\z}(z^+)])
\nonumber\\
&&-\frac{(x^+\!-\!y^+)^2}{4(k^+)^2}\;\U_{\{2\}}(x^+,y^+;[\hat{\z}(z^+)])
+ O\left(\left(\frac{(x^+\!-\!y^+)}{k^+}\, \partial_{\perp}^2\right)^3\right)\, ,
\eeq
with the next-to-eikonal term
\beq
\label{Semi_class_N2E_term}
\U_{\{1\}}(x^+,y^+;[\hat{\z}(z^+)]) &=& \P_+\; \U\left( x^+,y^+;[\hat{\z}(z^+)]\right)
\Bigg\{
\int_{y^+}^{x^+}\!\!\!\! dz^+_1\: \frac{(z_1^+\!-\!y^+)}{(x^+\!-\!y^+)}\; \delta^{ij}\, {\cal B}^{ij}(\mz_1)\nonumber\\
&&+2\int_{y^+}^{x^+}\!\!\!\! dz^+_1 \int_{z^+_1}^{x^+}\!\!\!\!  dz^+_2
\frac{(z^+_1\!-\!y^+)}{(x^+\!-\!y^+)}\, {\cal B}^i(\mz_2)\, {\cal B}^i(\mz_1)
\Bigg\}\, ,
\eeq
and the next-to-next-to-eikonal term
\beq
\label{Semi_class_N2N2E_term}
\U_{\{2\}}(x^+,y^+;[\hat{\z}(z^+)]) &=& \P_+\; \U\left( x^+,y^+;[\hat{\z}(z^+)]\right)
\Bigg\{
\frac{1}{2}\int_{y^+}^{x^+}\!\!\!\! dz_1^+\, \frac{(z_1^+\!-\!y^+)^2}{(x^+\!-\!y^+)^2}\,
\delta^{ij}\delta^{lm}\: {\cal B}^{ijlm}(\mz_1)\nonumber\\
& &\hspace{-4cm} + 2 \int_{y^+}^{x^+}\!\!\!\! dz^+_1  \int_{z^+_1}^{x^+}\!\!\!\!  dz^+_2\,
 \bigg[\frac{(z_2^+\!-\!y^+)(z_1^+\!-\!y^+)}{(x^+\!-\!y^+)^2}\,
 \left( \delta^{jl}\: {\cal B}^{ijl}(\mz_2)  {\cal B}^{i}(\mz_1) + \frac{1}{2} \delta^{ij}\delta^{lm}\: {\cal B}^{ij}(\mz_2)  {\cal B}^{lm}(\mz_1)
  \right)
\nonumber\\
&& \hspace{-2cm}
 +\frac{(z^+_1\!-\!y^+)^2}{(x^+\!-\!y^+)^2}\,\left(
{\cal B}^{i}(\mz_2)\: \delta^{jl}\, {\cal B}^{ijl}(\mz_1)
+{\cal B}^{ij}(\mz_2)\: {\cal B}^{ij}(\mz_1)\right)
 \bigg]
 \nonumber\\
&& \hspace{-4cm} +2\int_{y^+}^{x^+}\!\!\!\! dz^+_1\, \int_{z^+_1}^{x^+}\!\!\!\! dz^+_2\, \int_{z^+_2}^{x^+}\!\!\!\! dz^+_3\, \bigg[\frac{(z^+_3\!-\!y^+)(z^+_1\!-\!y^+)}{(x^+\!-\!y^+)^2} \delta^{ij} {\cal B}^{ij}(\mz_3)\; {\cal B}^l(\mz_2)\; {\cal B}^l(\mz_1)
\nonumber\\
&&\hspace{-3cm}
+2\frac{(z^+_1\!-\!y^+)^2}{(x^+\!-\!y^+)^2}{\cal B}^{i}(\mz_3)\; {\cal B}^{j}(\mz_2)\; {\cal B}^{ij}(\mz_1)
 +\frac{(z^+_2\!-\!y^+)(z^+_1\!-\!y^+)}{(x^+\!-\!y^+)^2}
\Big({\cal B}^{i}(\mz_3)\; \delta^{jl}{\cal B}^{jl}(\mz_2)\; {\cal B}^{i}(\mz_1)
\nonumber\\
&&\hspace{-3cm}
+{\cal B}^{i}(\mz_3)\; {\cal B}^{i}(\mz_2)\; \delta^{jl}{\cal B}^{jl}(\mz_1)
+2\; {\cal B}^{ij}(\mz_3)\; {\cal B}^{i}(\mz_2)\; {\cal B}^j(\mz_1)
+2\; {\cal B}^{i}(\mz_3)\; {\cal B}^{ij}(\mz_2)\; {\cal B}^{j}(\mz_1)
\Big)
\bigg]
\nonumber\\
&& \hspace{-4cm}
+4\int_{y^+}^{x^+}\!\!\!\! dz^+_1\, \int_{z^+_1}^{x^+}\!\!\!\! dz^+_2\,  \int_{z^+_2}^{x^+}\!\!\!\! dz^+_3\, \int_{z^+_3}^{x^+}\!\!\!\! dz^+_4
\bigg[ \frac{(z^+_3\!-\!y^+)(z^+_1\!-\!y^+)}{(x^+\!-\!y^+)^2}\;
{\cal B}^{i}(\mz_4)\;\; {\cal B}^i(\mz_3)\;\; {\cal B}^{j}(\mz_2)\;\; {\cal B}^{j}(\mz_1)
\nonumber\\
&&\hspace{-3cm}
+\frac{(z^+_2\!-\!y^+)(z^+_1\!-\!y^+)}{(x^+\!-\!y^+)^2}\;
 {\cal B}^{i}(\mz_4)\;\; {\cal B}^j(\mz_3)\, \Big( {\cal B}^i(\mz_2)\;\; {\cal B}^j(\mz_1)
+ {\cal B}^j(\mz_2)\;\; {\cal B}^i(\mz_1)
\Big)
\bigg]
\Bigg\}\, .
\eeq
Thanks to longitudinal boost invariance, the expansion \eqref{R_semiclass_exp} can be understood either as a large $k^+$ expansion or as a small $x^+\!-\!y^+$ expansion. The small parameter $(x^+\!-\!y^+)/k^+$ is dimensionful, but this is compensated by the appearance of two more transverse derivatives of the background field at each order in the expansion.



\subsection{Small angle expansion for the background propagator  \label{sec:small_angle_exp}}
So far, we have performed the large $k^+$ expansion of $\widetilde{\R}^{ab}_{\uk}(x^+,y^+;\y)$ at fixed $\k/k^+$, with the result given in eq. \eqref{R_semiclass_exp}. This is enough if one is interested in hard processes i.e. in the Bjorken limit. However, we are mostly interested in processes in the high-energy (or Regge) limit. Then, we have to re-expand the expression \eqref{R_semiclass_exp} in the small deflection angle limit $\k/k^+\rightarrow 0$. In eq. \eqref{R_semiclass_exp}, the quantity $\k/k^+$ appears through the trajectory $\hat{\z}(z^+)$ of the (decorated) Wilson lines, defined in eq. \eqref{zhat_cont}.

The term of zeroth order in eq. \eqref{R_semiclass_exp} is just the Wilson line $\U(x^+,y^+;[\hat{\z}(z^+)])$ along the trajectory $\hat{\z}(z^+)$. Its Taylor expansion with respect to $\k/k^+$ at second order is straightforward to perform, and gives
\beq
\U \left( x^+,y^+;[\hat{\z}(z^+)]\right)&=& \U(x^+,y^+;\y)+(x^+\!-\!y^+)\, \frac{\k^i}{k^+}\;\; \U^i_{[0,1]}(x^+,y^+;\y)
\nonumber\\
&& \hspace{-3cm}
+\frac{1}{2}(x^+\!-\!y^+)^2\, \frac{\k^i\k^j}{(k^+)^2}\;\; \U^{ij}_{[0,2]}(x^+,y^+;\y)
+ O\left(\left(\frac{(x^+\!-\!y^+)|\k|}{k^+}\, \partial_{\perp}\right)^3\right)\, ,
\eeq
where
\beq
\label{U01}
&&
\hspace{-0.9cm}
\U^i_{[0,1]}(x^+,y^+;\y)=\P_+\; \U(x^+,y^+;\y)\int_{y^+}^{x^+}\!\!\!\! dz_1^+\, \frac{(z_1^+\!-\!y^+)}{(x^+\!-\!y^+)} \; {\cal B}^i(z_1^+,\y)\, ,
\eeq
and\footnote{It is important, at this point, to make a clarification on the notation of the decorated Wilson lines. Here after, each decorated Wilson line appears with a subscript  $[\alpha, \beta]$ where both $\alpha$ and $\beta$ can take values from  0 to 2. In this subscript $\alpha$ stands for the order of expansion around the classical trajectory and $\beta$ stands for the order of small angle expansion. For example, $\U^i_{[0,1]}$ is the decorated Wilson line originating from the zeroth order expansion around the classical trajectory and the first order small angle expansion. We also would like to emphasise that $\U^i_{[0,1]}$ and $\U_{[1,0]}$ were called $\U^i_{(1)}$ and $\U_{(2)}$ respectively in Ref. \cite{Altinoluk:2014oxa}.}
\beq
\label{U02}
&&
\hspace{-0.9cm}
\U^{ij}_{[0,2]}(x^+,y^+;\y)=\P_+\; \U(x^+,y^+;\y)
\Bigg\{ \int_{y^+}^{x^+}\!\!\!\! dz_1^+\, \left(\frac{z_1^+\!-\!y^+}{x^+\!-\!y^+}\right)^2\; {\cal B}^{ij}(z_1^+,\y)\nonumber\\
&&\hspace{-0.7cm}
+2\int_{y^+}^{x^+}\!\!\!\! dz^+_1 \int_{z^+_1}^{x^+}\!\!\!\! dz^+_2\, \frac{(z^+_2\!-\!y^+)(z^+_1\!-\!y^+)}{(x^+\!-\!y^+)^2}\; {\cal B}^{i}(z^+_2,\y)\;\; {\cal B}^j(z_1^+,\y)\Bigg\}\, .
\eeq

In contrast, we need to Taylor-expand $\U_{\{1\}}(x^+,y^+;[\hat{\z}(z^+)])$ only to first order in $\k/k^+$ since it appears in the first order term in eq. \eqref{R_semiclass_exp}. According to eq. \eqref{Semi_class_N2E_term}, $\U_{\{1\}}(x^+,y^+;[\hat{\z}(z^+)])$ also depends on $\k/k^+$ only through the trajectory $\hat{\z}(z^+)$, which appears both in the Wilson line itself and in the decorations. Hence, one gets
\beq
\label{U1expanded}
&&\U_{\{1\}}(x^+,y^+;[\hat{\z}(z^+)])= \P_+\bigg\{\U(x^+,y^+;\y)
+(x^+\!-\!y^+)\frac{\k^l}{k^+}\;\, \U^l_{[0,1]}(x^+,y^+;\y) \bigg\}\nonumber\\
&&\times\bigg\{ \int_{y^+}^{x^+}\!\!\!\! dz^+\frac{(z^+\!-\!y^+)}{(x^+\!-\!y^+)}\Big[
\delta^{ij}\; {\cal B}^{ij}(z^+,\y) +(z^+\!-\!y^+)\frac{\k^i}{k^+}\, \delta^{jl}\; {\cal B}^{ijl}(z^+,\y)\Big]\nonumber\\
&&\hspace{0.5cm}+2\int_{y^+}^{x^+}\!\!\!\! dz_1^+ \int_{y^+}^{z_1^+}\!\!\!\! dz^+_2 \frac{(z^+_2\!-\!y^+)}{(x^+\!-\!y^+)}\bigg[
{\cal B}^{i}(z^+_1,\y)\;\; {\cal B}^{i}(z^+_2,\y)\nonumber\\
&&\hspace{0.5cm}+(z^+_1\!-\!y^+)\, \frac{\k^j}{k^+}\; {\cal B}^{ij}(z_1^+,\y)\;\; {\cal B}^{i}(z_2^+,\y)
+ (z^+_2\!-\!y^+)\, \frac{\k^j}{k^+}\;
{\cal B}^{i}(z^+_1,\y)\;\; {\cal B}^{ij}(z^+_2,\y)\bigg]
\bigg\}\nonumber\\
&&\hspace{3cm}
+ O\left(\left(\frac{(x^+\!-\!y^+)|\k|}{k^+}\, \partial_{\perp}\right)^2\right)\, .
\eeq
Eq. \eqref{U1expanded} can be organised and rewritten as
\beq
\label{U1final}
&&\hspace{-2cm}
\U_{\{1\}}(x^+,y^+;[\hat{\z}(z^+)])=\U_{[1,0]}(x^+,y^+;\y) +\frac{(x^+-y^+)\,\k^j}{(k^+)}\; \U^j_{[1,1]}(x^+,y^+;\y)
\nonumber\\
&& \hspace{3cm}
+ O\left(\left(\frac{(x^+\!-\!y^+)|\k|}{k^+}\, \partial_{\perp}\right)^2\right)\, ,
\eeq
where
\beq
\label{U10}
\U_{[1,0]}(x^+,y^+;\y)=\P_+\; \U(x^+,y^+;\y) \bigg\{\int_{y^+}^{x^+}\!\!\!\! dz_1^+\, \frac{(z_1^+\!-\!y^+)}{(x^+\!-\!y^+)}
\delta^{ij}\; {\cal B}^{ij}(z_1^+,\y)\nonumber\\
+2\int_{y^+}^{x^+}\!\!\!\! dz^+_1\, \int_{y^+}^{z^+_1}\!\!\!\! dz^+_2\, \frac{(z^+_2\!-\!y^+)}{(x^+\!-\!y^+)}\;
{\cal B}^{i}(z^+_1,\y)\;\; {\cal B}^{i}(z^+_2,\y)\bigg\}
\eeq
and
\beq
\label{U11}
&&\hspace{-0.5cm}
\U^i_{[1,1]}(x^+,y^+;\y)=\P_+\; \U(x^+,y^+;\y)\Bigg\{\int_{y^+}^{x^+}\!\!\!\! dz_1^+\, \left(\frac{z_1^+\!-\!y^+}{x^+\!-\!y^+}\right)^2
\delta^{jl}\; {\cal B}^{ijl}(z_1^+,\y)
\nonumber\\
&&\hspace{-0.5cm}
+2\int_{y^+}^{x^+}\!\!\!\! dz^+_1\, \int_{z^+_1}^{x^+}\!\!\!\! dz^+_2\,
\Bigg[\left(\frac{z^+_1\!-\!y^+}{x^+\!-\!y^+}\right)^2\;
{\cal B}^{j}(z^+_2,\y)\;\; {\cal B}^{ij}(z^+_1,\y)
\nonumber\\
&& \hspace{2cm}
+\frac{(z^+_1\!-\!y^+)(z^+_2\!-\!y^+)}{(x^+\!-\!y^+)^2}\Bigg(
{\cal B}^{ij}(z^+_2,\y)\;\; {\cal B}^{j}(z^+_1,\y)
\nonumber\\
&& \hspace{3cm}
+\frac{1}{2}\, {\cal B}^{i}(z^+_2,\y)\;\; \delta^{jl}\; {\cal B}^{jl}(z^+_1,\y)
+\frac{1}{2}\delta^{jl}\; {\cal B}^{jl}(z^+_2,\y)\;\; {\cal B}^{i}(z^+_1,\y)
\Bigg)\Bigg]\nonumber\\
&& \hspace{-0.5cm}
+2\int_{y^+}^{x^+}\!\!\!\! dz^+_1\, \int_{z^+_1}^{x^+}\!\!\!\! dz^+_2\, \int_{z^+_2}^{x^+}\!\!\!\! dz^+_3\,
\bigg[\frac{(z^+_3\!-\!y^+)(z^+_1\!-\!y^+)}{(x^+\!-\!y^+)^2}
{\cal B}^i(z^+_3,\y)\;\; {\cal B}^j(z^+_2,\y)\;\; {\cal B}^{j}(z^+_1,\y)
\nonumber\\
&& \hspace{-0.5cm}
+\frac{(z^+_2\!-\!y^+)(z^+_1\!-\!y^+)}{(x^+\!-\!y^+)^2}
{\cal B}^j(z^+_3,\y)\;
\Big(
{\cal B}^i(z^+_2,\y)\; {\cal B}^{j}(z^+_1,\y)
+{\cal B}^j(z^+_2,\y)\; {\cal B}^{i}(z^+_1,\y)
\Big)\bigg]\Bigg\}\, .
\eeq

For the last term in eq. \eqref{R_semiclass_exp}, which is already of next-to-next-to-eikonal order, we can drop completely the $\k/k^+$ dependence, or equivalently make the replacement $\hat{\z}(z^+)\mapsto \y$. We thus write
\beq
\label{U2_exp}
&&\hspace{-2cm}
\U_{\{2\}}(x^+,y^+;[\hat{\z}(z^+)])=\U_{[2,0]}(x^+,y^+;\y)
+ O\left(\frac{(x^+\!-\!y^+)|\k|}{k^+}\, \partial_{\perp}\right)\, ,
\eeq
where
\beq
\label{U20}
&&\hspace{-2cm}
\U_{[2,0]}(x^+,y^+;\y)\equiv \U_{\{2\}}(x^+,y^+;[\y])
\, ,
\eeq
which can be read from eq. \eqref{Semi_class_N2N2E_term}.

All in all, after both the semi-classical and the small angle expansions, the medium modification factor $\widetilde{\R}^{ab}_{\uk}(x^+,y^+;\y)$ for the propagator reads, at next-to-next-to-eikonal accuracy,
\beq
\label{Rtildefinal}
\hspace{-1cm}
\widetilde{\R}_{\uk}(x^+,y^+;\y)&=&\U(x^+,y^+;\y) +\frac{(x^+\!-\!y^+)\k^i}{k^+}\; \U^i_{[0,1]}(x^+,y^+;\y)
\nonumber\\
&+&
i\frac{(x^+\!-\!y^+)}{2k^+}\; \U_{[1,0]}(x^+,y^+;\y)
+\frac{(x^+\!-\!y^+)^2\k^i\k^j}{(k^+)^2}\; \U^{ij}_{[0,2]}(x^+,y^+;\y)
\nonumber\\
&+&
i\frac{(x^+\!-\!y^+)^2\k^i}{2(k^+)^2}\; \U^i_{[1,1]}(x^+,y^+;\y) -\frac{(x^+\!-\!y^+)^2}{4(k^+)^2}\; \U_{[2,0]}(x^+,y^+;\y)
\nonumber\\
& & + O\left(\left(\frac{(x^+\!-\!y^+)|\k|}{k^+}\, \partial_{\perp}\right)^3\right)
    + O\left(\left(\frac{(x^+\!-\!y^+)}{k^+}\, \partial_{\perp}^2\right)^3\right),
\eeq
where the various decorated Wilson lines are defined in Eqs. \eqref{U01},  \eqref{U02}, \eqref{U10}, \eqref{U11}, and \eqref{U20}.

\section{Single inclusive gluon production in pA collisions}
\label{spectra}

The eikonal expansion performed in the previous section at the level of the gluon background propagator can be applied to single inclusive gluon production in pA collisions at high energy. This process can be treated in the CGC effective theory assuming that the gluon is produced at mid-rapidity.

Within the CGC framework, following the same idea developed in section 2, we describe the highly boosted left-moving nucleus by the background field given in eq. \eqref{Background_field_def} with a finite longitudinal support from $x^+=0$ to $x^+=L^+$. The right-moving proton, on the other hand, is considered to be dilute and described by a classical color current
\beq
j^{\mu}(x)=\delta^{\mu -}j^+_a(x)
\label{jmu}
\eeq
which is localized at $x^-=0$.

Let us consider a proton-nucleus collision with some impact parameter $\B$, taking the centre of the nucleus as the reference point for the transverse plane. Then an arbitrary point $\x$ in the transverse plane is at a distance $|\x-\B|$ from the center of the proton and at a distance $|\x|$ from the center of the nucleus. In this setup, the classical color current $j^+(x)$ is written as
\beq
j^{+}_a(x)=\delta(x^-)\; \U^{ab}(x^+,-\infty;\x)\;\;  \rho^{b}(\x\!-\!\B),
\eeq
where $\rho^b$ is the transverse color charge density inside the proton before the interaction with the nucleus, and $\U^{ab}(x^+,-\infty;\x)$ is the Wilson line implementing the color precession of these color charges in the background field $\A^-_a(x^+,\x)$ of the nucleus.

The single inclusive gluon production cross-section for a pA collision is given by
\beq
(2\pi)^3\, (2k^+)\, \frac{d\sigma}{dk^+\, d^2\k} = \int d^2\B \sum_{\lambda\, \textrm{phys.}}   \left\langle \left\langle  \left| {\cal M}^a_{\lambda}(\underline{k}, \B)\right|^2  \right\rangle_{p} \right\rangle_{A}\, ,
\label{Def_Sigma}
\eeq
where $\lambda$,  $a$ and $\underline{k}=(k^+,\k)$ are the polarization, color and momentum of the produced gluon. At leading order in the coupling $g$, the gluon production amplitude, ${\cal M}^a_{\lambda}$, is given by the LSZ-type reduction formula
\beq
{\cal M}^a_{\lambda}(\underline{k}, \B)&=& \varepsilon_{\lambda}^{\mu *}  \int d^4x\; e^{i k\cdot x}\; \Box_{x} A_{\mu}^a(x)\ ,
\label{LSZ}
\eeq
where $A^a_{\mu}(x)$ is the retarded classical field \cite{Gelis:2006yv}. Since we are interested in dilute-dense scattering, the background field ${\cal A}^-_a(x^+,\x)$ that is describing the dense nucleus is $O(1/g)$ and the color charge current $j^+_a(x)$ that is describing the dilute  proton is $O(g)$. The retarded classical field $A^a_{\mu}(x)$ appearing in the reduction formula contains both the background field ${\cal A}^-_a(x^+,\x)$ and small perturbations on top of it, generated by the color current $j^+_a(x)$. However, since in the light cone gauge $A^+=0$ the physical polarization vectors satisfy $\varepsilon_{\lambda}^{+ *}=0$, only the transverse components of the field perturbation due to $j^+_a(x)$ contributes to eq. \eqref{LSZ}, which can be written at leading order in $g$ as
\beq
{\cal M}^a_{\lambda}(\underline{k}, \B)&=&  \varepsilon_{\lambda}^{i *}\:  \lim_{x^+\rightarrow +\infty} e^{i k^- x^+} \int d^2\x\; e^{-i \k\cdot \x}  \int d^4 y\;
{\cal G}^{i-}_{k^+}(\underline{x};\underline{y})_{ab}\;\; j^{+}_b(y)\, .
\label{M1}
\eeq

The gluon production amplitude ${\cal M}^a_{\lambda}(\underline{k}, \B)$  can equivalently be written in momentum space as
\beq
{\cal M}^a_{\lambda}(\underline{k}, \B)&=& \int \frac{d^2 \q}{(2\pi)^2}\; e^{-i \q\cdot\B}\;\;    \overline{{\cal M}}^{a b}_{\lambda}(\underline{k}, \q)\;\; \tilde{\rho}^{b}(\q)\,
\label{def_reduced_Ampl}
\eeq
by defining a  gluon-nucleus reduced amplitude $\overline{{\cal M}}^{a b}_{\lambda}(\underline{k}, \q)$, and $\tilde{\rho}(\q)$ through the Fourier transform
\beq
\rho^{a}(\y-\B)= \int \frac{d^2 \q}{(2\pi)^2}\; e^{i \q\cdot(\y-\B)}\; \tilde{\rho}^{a}(\q)\, .
\label{Fourier_rho}
\eeq
It is clear from eq. \eqref{Def_Sigma} and eq. \eqref{def_reduced_Ampl} that one needs to average over the color charge densities at equal momenta, $\left\langle \tilde{\rho}^a(\q)\tilde{\rho}^b(\q)\right\rangle_p$, in the calculation of the single inclusive gluon production cross-section, due to the integration over the impact parameter $\B$. It has been shown and discussed in detail in Ref. \cite{Altinoluk:2014oxa} that this correlator is indeed related to the unintegrated gluon distribution $\varphi_p(\q)$\footnote{Here, we indicate only the dependence of $\varphi_p$ on the transverse momentum of the gluon, both for simplicity and by consistency with the fact that our calculation is only a leading order one in the coupling. On the one hand, due to the boost-invariance of the classical Weizs\"acker-Williams gluon field sourced by the current $j^{\mu}(x)$, $\varphi_p(\q)$ does not depend on the $k^+$ of the gluon at this order. On the other hand, when including radiative corrections, $\varphi_p(\q)$ would acquire dependence on a factorization scale regulating the soft divergence. In order to resum optimally the small-$x$ leading logs for the projectile, one should then take for the factorization scale a value related to the $k^+$ of the gluon. This situation is analog to the case of collinear factorization, where the parton densities acquire a dependence on the hard scale through the choice of renormalization scale, which is absent in a leading order calculation in the parton model.
} in the projectile via
\beq
\left\langle \tilde{\rho}^{a}(\q)^{*}\: \tilde{\rho}^{b}(\q) \right\rangle_{p}&=&
\frac{\delta^{ab}}{N_c^2\!-\!1}\;\;  \frac{(2\pi)^3}{2}\;\; \q^2\;\; \varphi_p(\q)
\label{rhorho_vs_ugd}\, .
\eeq

\subsection{Corrections to the eikonal limit at the amplitude level }
The gluon-nucleus reduced amplitude $\overline{\cal M}^{ab}_{\lambda}(\underline{k},\q)$ can be decomposed into three contributions as
\beq
\overline{\cal M}^{ab}_{\lambda}(\underline{k},\q)=\overline{\cal M}^{ab}_{bef,\lambda}(\uk,\q)+\overline{\cal M}^{ab}_{in,\lambda}(\uk,\q)+\overline{{\cal M}}^{a b}_{aft,\lambda}(\underline{k}, \q),
\eeq
in which the gluon is radiated by the color current respectively before, during or after the interaction with the background field $\A^-_a(x^+,\x)$ . Each contribution can be written explicitly as
\beq
&&\overline{\cal M}^{ab}_{bef,\lambda}(\uk,\q)=\epsilon^{i*}_{\lambda}e^{ik^-L^+}i\int d^2\y\;e^{i\q\cdot\y} \;(-2)\;\frac{\q^i}{\q^2}\int d^2\z e^{-i\k\cdot\z}\G_{k^+}^{ab}(L^+,\z;0,\y),
\label{befred}
\\\nonumber\\
%
&&\overline{\cal M}^{ab}_{in,\lambda}(\uk,\q)=\epsilon^{i*}_{\lambda}e^{ik^-L^+}i\int d^2\y\; e^{i\q\cdot\y}\frac{1}{k^+}\int_0^{L^+}dy^+\Bigg\{\del_{\y^i}\bigg[\int d^2\z\, e^{-i\k\cdot\z}\, \G_{k^+}^{ac}(L^+,\z;\uy)\bigg]\Bigg\}\nonumber\\
&&\hspace{8.3cm}\times \; \U^{cb}(y^+,0;\y) \, ,
\label{inred}
\\\nonumber\\
%
&&\overline{{\cal M}}^{a b}_{aft,\lambda}(\underline{k}, \q)=  \varepsilon_{\lambda}^{i *}\: e^{i k^- L^+}\; i  \int d^2\y\;\; e^{i (\q-\k)\cdot \y}\;\;\; 2\, \frac{\k^i}{\k^2}
\;\;   \U^{ab}(L^+,0,\y)\, .
\label{aftred}
\eeq

The "$aft$" contribution to the total gluon-nucleus reduced amplitude does not involve the background propagator $ \G^{ab}_{k^+}(\ux,\uy)$, thus it can be kept as it is written in eq. \eqref{aftred}. On the other hand, the "$bef$" and "$in$" contributions to the total reduced amplitude involves the background propagator whose eikonal expansion has been performed in section 2. Thus, one can use the expanded expression of the background propagator $ \G^{ab}_{k^+}(\ux,\uy)$ in eqs.\eqref{befred} and \eqref{inred} in order to calculate the corrections to the eikonal limit of the "$bef$" and "$in$" contributions to the total gluon-nucleus reduced amplitude.

Let us first consider the "$bef$" contribution. By using eq. \eqref{Rtilde_def}, $\overline{\cal M}^{ab}_{bef,\lambda}(\uk,\q)$ can be written in terms of the medium modification factor  $\widetilde{\R}_{\uk}(x^+,y^+;\y)$ can be written as
\beq
\overline{\cal M}^{ab}_{bef,\lambda}(\uk,\q)=\epsilon^{i*}_{\lambda}\int d^2\y\; e^{i(\q-\k)\cdot\y}\; (-2)\;\frac{\q^i}{\q^2}\;\widetilde{\R}^{ab}_{\uk}(L^+,0;\y)\, .
\label{befredR}
\eeq
One can now use the expression of the expanded $\widetilde{\R}^{ab}_{\uk}(L^+,0;\y)$ that is given in eq. \eqref{Rtildefinal} in eq. \eqref{befredR} to write the "$bef$" contribution to the total gluon-nucleus reduced amplitude at next-to-next-to-eikonal accuracy as
\beq
&&\overline{M}^{ab}_{bef,\lambda}(\uk,\q)=i\int d^2\y\; e^{i(\q-\k)\cdot\y}\; (-2)\;\frac{(\epsilon^{*}_\lambda\cdot\q)}{\q^2}\bigg\{\U(L^+,0;\y)+\frac{L^+}{k^+} \;\k^j\;\U^j_{[0,1]}(L^+,0;\y)\nonumber\\
&&+i\frac{L^+}{2k^+}\; \U_{[1,0]}(L^+,0;\y)+\left(\frac{L^+}{k^+}\right)^2\k^i\k^j\;\U^{ij}_{[0,2]}(L^+,0;\y)+\frac{i}{2}\left(\frac{L^+}{k^+}\right)^2\k^j\;\U^{j}_{[1,1]}(L^+,0;\y)\nonumber\\
&&-\frac{1}{4}\left(\frac{L^+}{k^+}\right)^2\U_{[2,0]}(L^+,0;\y)\bigg\}^{ab}\, .
\label{bef_red_exp}
\eeq

The remaining contribution that involves the background propagator $\G^{ab}_{k^+}(\ux,\uy)$ is the "$in$" term $\overline{M}^{ab}_{in,\lambda}(\uk,\q)$, given in eq. \eqref{inred}. Inserting eq. \eqref{Rtilde_def}, into eq. \eqref{inred}, one gets
\beq
&&\overline{M}^{ab}_{in,\lambda}(\uk,\q)=\epsilon^{i*}_{\lambda}\;i\int d^2\y\; e^{-i(\k-\q)\cdot\y}\frac{1}{k^+}\int_0^{L^+}\!\!\!\!  dy^+\; e^{iy^+k^-}\; \bigg[\left( \del_{\y^i}-i\k^i\right)\widetilde{\R}^{ac}_{\uk}(L^+,y^+;\y)\bigg]\nonumber\\
&&\hspace{8.3cm}\times\; \U^{cb}(y^+,0;\y)\, .
\label{In_Red_wR}
\eeq
The presence of an explicit factor of $1/k^+$ in eq. \eqref{In_Red_wR} suggests that it is enough to keep the terms of $O\left(1\right)$ and of $O\left(L^+/k^+\right)$ in the expansion of the medium modification factor $\widetilde{\R}^{ab}_{\uk}(L^+,0;\y)$ in order to get the expression of the $\overline{M}^{ab}_{in,\lambda}(\uk,\q)$ at next-to-next-to-eikonal accuracy. Keeping this in mind and using the expanded expression of the medium modification factor $\widetilde{\R}^{ab}_{\uk}(L^+,0;\y)$ given in eq. \eqref{Rtildefinal}, $\overline{M}^{ab}_{in,\lambda}(\uk,\q)$ can be written as
\beq
&&\hspace{-1cm}\overline{M}^{ab}_{in,\lambda}(\uk,\q)=\epsilon^{i*}_{\lambda}\;i\int d^2\y\; e^{i(\q-\k)\cdot\y}\frac{1}{k^+}\int_0^{L^+}\!\!\!\!  dy^+\; e^{iy^+k^-}\left( \del_{\y^i}-i\k^i\right) \bigg[ \U(L^+,y^+;\y)\nonumber\\
&&\hspace{-1cm}+(L^+\!-\!y^+)\frac{\k^j}{k^+}{\U^{j}}_{[0,1]}(L^+,y^+;\y) +\frac{i}{2k^+}(L^+\!-\!y^+)\U_{[1,0]}(L^+,y^+;\y)\bigg]^{ac}\U^{cb}(y^+,0;\y)\, .
\label{in_red_exp}
\eeq

Note that eq. \eqref{in_red_exp} contains terms with the transverse derivative acting on both the usual Wilson line $\U^{ab}(L^+,y^+; \y)$ and also on the {\it decorated Wilson lines} $\U^{j \; ab}_{[0,1]}(L^+,y^+;\y)$ and $\U^{ab}_{[1,0]}(L^+,y^+;\y)$. In order to have an explicit and complete expression for the "$in$" contribution to the gluon-nucleus reduced amplitude at next-to-next-to-eikonal accuracy, one should calculate these terms explicitly. Transverse derivative of the usual Wilson $\U^{ab}(L^+,y^+; \y)$ line has been calculated in Ref. \cite{Altinoluk:2014oxa} and the result reads
\beq
\del_{\y^i}\U^{ab}(L^+,y^+; \y)=
\int_{y^+}^{L^+}dz^+ \left[ \U(L^+,z^+;\y)\left( igT\cdot\del_{\y^i}\A^-(z^+,\y)\right)\U(z^+,0;\y)
\right]^{ab}\, .
\label{delU}
\eeq
On the other hand, the transverse derivative of the {\it decorated Wilson lines} have not appeared  at next-to-eikonal accuracy, thus they have not been calculated in \cite{Altinoluk:2014oxa}. Given the fact that the decorated Wilson lines are composed of the usual Wilson lines and field insertions between these usual Wilson lines, it is straightforward to calculate the transverse derivative of the decorated Wilson lines. When the derivative is acting on the field insertion, it simply increases the number of derivatives acting on the background field $\A^-(z^+,\y)$. When it is acting on the usual Wilson line, it introduces an extra field insertion as shown in eq. \eqref{delU}. Then, one obtains
\beq
\del_{\y^i}\U^{j}_{[0,1]}(L^+,y^+;\y)&=&\P_+\,\U(L^+,y^+;\y)\int_{y^+}^{L^+}\!\!\!\! dz^+\, \frac{z^+\!-\!y^+}{L^+\!-\!y^+}\; {\cal B}^{ij}(z^+,\y)\nonumber\\
&&\hspace{-2cm}+\;\P_+\,\U(L^+,y^+;\y)\int_{y^+}^{L^+}\!\!\!\! dw^+\; {\cal B}^{i}(w^+,\y) \int_{y^+}^{L^+}\!\!\!\! dz^+\, \frac{z^+\!-\!y^+}{L^+\!-\!y^+}\; {\cal B}^{j}(z^+,\y)\, .
\eeq
Here we only present the transverse derivative of one of the decorated Wilson lines but the same arguments hold for the other one and it can also be calculated straightforwardly. All in all, the "$in$" contribution to the gluon-nucleus reduced amplitude $\overline{M}^{ab}_{in,\lambda}(\uk,\q)$ at next-to-next-to-eikonal accuracy can be written as
\beq
&&\hspace{-0.4cm}\overline{M}^{ab}_{in,\lambda}(\uk,\q)=\epsilon^{i*}_{\lambda}\;i\int d^2\y\; e^{i(\q-\k)\cdot\y}\bigg\{
2\,\frac{\k^i}{\k^2}\left(1-e^{iL^+k^-}\right)\U^{ab}(L^+,0;\y) \nonumber\\
&&\hspace{-0.4cm}+\frac{L^+}{k^+}\; \U^{i\; ab}_{[0,1]}(L^+,0;\y)+\left(\frac{L^+}{k^+}\right)^2 \bigg[\frac{i}{4}\left(\k^2\delta^{ij}\!-\!2\k^i\k^j\right) \U^{j}_{({\rm A})}(L^+,0;\y)
 \nonumber\\
&&\hspace{4cm} +\frac{\k^j}{4}\U^{ij}_{(\rm B)}(L^+,0;\y) +\frac{i}{4}\U^{i}_{(\rm C)}(L^+,0;\y)\bigg]^{ab}\bigg\}\, .
\label{in_red_exp2}
\eeq
Here, $\U^{j}_{({\rm A})}(L^+,0;\y)$, $\U^{ij}_{(\rm B)}(L^+,0;\y)$ and $\U^{i}_{(\rm C)}(L^+,0;\y)$ are the new decorated Wilson line structures that are defined as
follows:
\beq
\U^{i\; ab}_{({\rm A})}(L^+,0;\y)&=&\P_+\; \U^{ab}(L^+,0;\y) \int_0^{L^+}\!\!\!\! dz^+\, \left(\frac{z^+}{L^+}\right)^2\; {\cal B}^{i}(z^+,\y),
\eeq
\beq
&&\U^{ij\; ab}_{({\rm B})}(L^+,0;\y)=\P_+\; \U^{ab}(L^+,0;\y) \bigg\{ \Big[ \delta^{ij}\delta^{lm} +\delta^{il}\delta^{jm}+\delta^{im}\delta^{jl}\Big] \int_0^{L^+}\!\!\!\! dz^+\, \left(\frac{z^+}{L^+}\right)^2\; {\cal B}^{lm}(z^+,\y)\nonumber\\
&&+2\Big[ \delta^{ij}\delta^{lm} +\delta^{il}\delta^{jm}+\delta^{im}\delta^{jl}\Big] \int_0^{L^+}\!\!\!\! dz_1^+\, \int_{z^+_1}^{L^+}\!\!\!\! dz^+_2\, \left(\frac{z^+_1}{L^+}\right)^2\; {\cal B}^l(z^+_2,\y)\; {\cal B}^{m}(z^+_1,\y)\nonumber\\
&&+4 \int_0^{L^+}\!\!\!\! dz_1^+\, \int_{z^+_1}^{L^+}\!\!\!\! dz^+_2\, \frac{z^+_1(z^+_2\!-\!z^+_1)}{(L^+)^2}\; {\cal B}^j(z^+_2,\y)\; {\cal B}^{i}(z^+_1,\y)\bigg\},
\eeq
\beq
&&\U^{i\; ab}_{({\rm C})}(L^+,0;\y)=\P_+\; \U^{ab}(L^+,0;\y) \bigg\{ \int_0^{L^+}\!\!\!\! dz^+\, \left(\frac{z^+}{L^+}\right)^2\; \delta^{jl}\; {\cal B}^{ijl}(z^+,\y)\nonumber\\
&&+\int_0^{L^+}\!\!\!\! dv^+\, \int_0^{L^+}\!\!\!\! dz^+\, \bigg[\theta(v^+\!-\!z^+)\, \left(\frac{z^+}{L^+}\right)^2 +\theta(z^+\!-\!v^+)\, \frac{v^+}{(L^+)^2} \left(2\, z^+\!-\!v^+\right)\bigg]\nonumber\\
&&\hspace{3.5cm}\times\; \delta^{jl}\, {\cal B}^{jl}(z^+,\y)\; {\cal B}^{i}(v^+,\y)\nonumber\\
&&+2\int_0^{L^+}\!\!\!\! dz^+_2\, \int_0^{z^+_2}\!\!\!\! dz^+_1\,  \left(\frac{z^+_1}{L^+}\right)^2 \bigg[ {\cal B}^{ij}(z^+_2,\y)\; {\cal B}^{j}(z^+_1,\y)
+{\cal B}^{j}(z^+_2,\y)\; {\cal B}^{ij}(z^+_1,\y)\bigg]\nonumber\\
&&+2\int_0^{L^+}\!\!\!\! dv^+\, \int_0^{L^+}\!\!\!\! dz^+_2\, \int_0^{z^+_2}\!\!\!\! dz^+_1\, \bigg[\theta(v^+\!-\!z^+_1) \left(\frac{z^+_1}{L^+}\right)^2 +\theta(z^+_1\!-\!v^+) \frac{v^+}{(L^+)^2} \left(2\, z^+_1\!-\!v^+\right)\bigg]\nonumber\\
&&\hspace{5.3cm}\times\; {\cal B}^i(v^+,\y)\; {\cal B}^j(z^+_2,\y)\; {\cal B}^j(z^+_1,\y)\bigg\}\, .
\eeq

Finally, by using eqs. \eqref{aftred}, \eqref{bef_red_exp} and \eqref{in_red_exp2} the total gluon-nucleus reduced amplitude $\overline{\mathcal{M}}^{ab}_{\lambda} \left(\underline{k};\q\right)$ at next-next-to-eikonal accuracy can be written as
\beq
&& \overline{\mathcal{M}}^{ab}_{\lambda} \left(\underline{k};\q\right)=i \, \varepsilon^{i*}_{\lambda}\int d^2\y \, e^{i\y\cdot(\q-\k)}
\Bigg\{
2\left(\frac{\k^i}{\k^2}-\frac{\q^i}{\q^2}\right) \mathcal{U}(L^+,0;\y)\nonumber\\
&&+\left( \frac{L^+}{k^+}\right)\bigg[
\left(\delta^{ij}-2\, \frac{\q^i\k^j}{\q^2}\right) \mathcal{U}^j_{[0,1]}(L^+,0;\y)
 -i\frac{\q^i}{\q^2}\; \mathcal{U}_{[1,0]}(L^+,0;\y)\bigg]\nonumber\\
 &&+\left(\frac{L^+}{k^+}\right)^2
\Bigg[
-2\frac{\q^i}{\q^2}\k^j\k^l\; \mathcal{U}^{jl}_{[0,2]}(L^+,0;\y)
-i\frac{\q^i\k^j}{\q^2}\; \mathcal{U}^j_{[1,1]}(L^+,0;\y)
+\frac{1}{2}\frac{\q^i}{\q^2}\; \mathcal{U}_{[2,0]} (L^+,0;\y)\nonumber \\
&& +\frac{i}{4}\left(\k^2\delta^{ij}-2\k^i\k^j\right)\, \mathcal{U}^j_{(\rm A)}(L^+,0;\y)
+\frac{\k^j}{4}\; \mathcal{U}^{ij}_{(\rm B)}(L^+,0;\y)
+\frac{i}{4}\; \mathcal{U}^i_{(\rm C)}(L^+,0;\y)
\Bigg]
\Bigg\}^{ab}\, .
\label{final_red_total}
\eeq

\subsection{Gluon production cross section beyond eikonal accuracy}

The single inclusive gluon production cross section and light-front helicity asymmetry of the produced gluon are the two observables that have been discussed in detail in Ref. \cite{Altinoluk:2014oxa} at next-to-eikonal accuracy. For the single inclusive gluon production cross section, it has been shown that the next-to-eikonal terms vanish and the well known result - the strict eikonal limit of the $k_{\perp}$-factorized formula - has been obtained. On the other hand, the strict eikonal terms for the light-front helicity asymmetry have been shown to vanish, leaving the next-to-eikonal contributions as the leading terms for this particular observable.

We have calculated the gluon-nucleus reduced amplitude $\overline{\mathcal{M}}^{ab}_{\lambda} \left(\underline{k};\q\right)$ at next-to-next-to-eikonal accuracy in the previous section. In this section, our aim is to use this result to calculate these two observables at next-to-next-to-eikonal accuracy.

The single inclusive gluon cross-section, eq. \eqref{Def_Sigma}, can be written in terms of the gluon-nucleus reduced amplitude $\overline{\mathcal{M}}^{ab}_{\lambda} \left(\underline{k};\q\right)$ as
\beq
\hspace{-0.2cm}
(2\pi)^3\, (2k^+)\, \frac{d\sigma}{dk^+\, d^2\k} = \int\frac{d^2\q}{(2\pi)^2}\left\langle \tilde\rho^c(\q)^*\tilde\rho^b(\q) \right\rangle_p\sum_{\lambda}\left\langle \left(\overline{\mathcal{M}}^{ac}_{\lambda}(\uk,\q)\right)^\dagger \overline{\mathcal{M}}^{ab}_{\lambda}(\uk,\q)\right\rangle_{A}\, .
\eeq
By using the relation, given in eq. \eqref{rhorho_vs_ugd}, between the correlator of two charge densities in the projectile $\left\langle \tilde\rho^c(\q)^*\tilde\rho^b(\q) \right\rangle_p$ and the unintegrated gluon distribution of the projectile  $\varphi_p(\q)$, the single inclusive gluon cross-section can be written as
\beq
k^+ \frac{d\sigma}{dk^+\, d^2\k}= \int\frac{d^2\q}{(2\pi)^2} \, \varphi_p(\q) \, \frac{\q^2}{4} \, \frac{1}{N^2_c-1}\sum_{\lambda}\left\langle \overline{\mathcal{M}}^{ab}_{\lambda}(\uk,\q)^\dagger \overline{\mathcal{M}}^{ab}_{\lambda}(\uk,\q)\right\rangle_{A} \; .
\label{SIGCS}
\eeq

The second observable that we consider is the light-front helicity asymmetry of the produced gluon. The calculation of this asymmetry is almost identical to single inclusive gluon production cross section, except that instead of summing over the helicity $\lambda\pm1$ of the produced gluon, one takes the difference between the $\lambda=+1$ and $\lambda=-1$ contributions i.e.
\beq
& & \hspace{-1cm} k^+\, \frac{d\sigma^+}{dk^+\, d^2\k}-k^+\, \frac{d\sigma^-}{dk^+\, d^2\k}=\int \frac{d^2 \q}{(2\pi)^2}\;
 \varphi_p(\q)\;  \nonumber\\
 && \hspace{4cm} \times\: \frac{ \q^2}{4}\;
\frac{1}{N_c^2\!-\!1}
\sum_{\lambda\, \textrm{phys.}} \lambda\; \left\langle  \overline{{\cal M}}^{a b}_{\lambda}(\underline{k}, \q)^{\dag}\;   \overline{{\cal M}}^{a b}_{\lambda}(\underline{k}, \q)  \right\rangle_{A} .
\label{gluon_asym_cross_section}
\eeq

In the calculation of both single inclusive gluon cross-section eq. \eqref{SIGCS} and light-front helicity asymmetry eq. \eqref{gluon_asym_cross_section}, one needs to take the square of the gluon-nucleus reduced amplitude $\overline{{\cal M}}^{a b}_{\lambda}(\underline{k}, \q)$. At the squared amplitude level, we keep the $\lambda$ dependence explicit so that we can apply the result to both observables in eqs. \eqref{SIGCS} and \eqref{gluon_asym_cross_section}. Keeping this in mind, the square of the amplitude can be expanded as
\beq
\overline{\mathcal{M}}^{ab}_{\lambda}(\uk,\q)^\dagger \overline{\mathcal{M}}^{ab}_{\lambda}(\uk,\q)&=& \left. \overline{\mathcal{M}}^{ab}_{\lambda}(\uk,\q)^\dagger \overline{\mathcal{M}}^{ab}_{\lambda}(\uk,\q)\right|_{\rm E}+\left. \overline{\mathcal{M}}^{ab}_{\lambda}(\uk,\q)^\dagger \overline{\mathcal{M}}^{ab}_{\lambda}(\uk,\q)\right|_{\rm NE}\nonumber\\
&&+\left. \overline{\mathcal{M}}^{ab}_{\lambda}(\uk,\q)^\dagger \overline{\mathcal{M}}^{ab}_{\lambda}(\uk,\q)\right|_{\rm NNE}+\cdots ,
\label{decomposedM}
\eeq
where the terms on the right hand side of eq. \eqref{decomposedM} stand for the eikonal terms, next-to-eikonal terms and next-to-next-to-eikonal terms respectively, and the dots for higher order terms.

By using the expression for the gluon-nucleus reduced amplitude $\overline{{\cal M}}^{a b}_{\lambda}(\underline{k}, \q)$ given in eq. \eqref{final_red_total} each contribution to the squared amplitude can be written explicitly. The eikonal contribution reads
\beq
\left. \overline{\mathcal{M}}^{ab}_{\lambda}(\uk,\q)^\dagger \overline{\mathcal{M}}^{ab}_{\lambda}(\uk,\q)\right|_{\rm E}&=&\varepsilon^{i*}_{\lambda}\varepsilon^{j}_{\lambda}\int d^2\y \int d^2\y^\prime \, e^{i(\y-\y^\prime)\cdot(\q-\k)}\nonumber\\
&&\times
4\;\C^i(\k,\q)\;\C^j(\k,\q) \tr\left[\mathcal{U}^\dagger(\y^\prime)\mathcal{U}(\y)\right]\, .
\label{M2E}
\eeq
Note that we have dropped the longitudinal coordinate dependence of the Wilson lines for convenience but one should keep in mind that these Wilson lines run from $0$ to $L^+$ in the $x^+$ direction i.e. $\U(\y)\equiv\U(L^+,0; \y)$. We will use this notation for the rest of the paper. Moreover, we have introduced the shorthand notation $\C^i(\k,\q)$ for the coefficient of the Wilson lines which is defined as
\beq
\C^i(\k,\q)&=&\left(\frac{\k^i}{\k^2}-\frac{\q^i}{\q^2}\right)\, .
\label{coefficient1}
\eeq
Similarly, the next-to-eikonal contribution reads
\beq
\left. \overline{\mathcal{M}}^{ab}_{\lambda}(\uk,\q)^\dagger \overline{\mathcal{M}}^{ab}_{\lambda}(\uk,\q)\right|_{\rm NE}&=&\varepsilon^{i*}_{\lambda}\varepsilon^{j}_{\lambda}\int d^2\y \int d^2\y^\prime \, e^{i(\y-\y^\prime)\cdot(\q-\k)} \;  2\left(\frac{L^+}{k^+}\right)     \nonumber\\
&&\hspace{-3.8cm}\times
\Bigg[
\C^j(\k,\q)\; \tC^{l i}(\k,\q)
 \tr \left[\mathcal{U}^\dagger(\y^\prime)\mathcal{U}^l_{[0,1]}(\y)\right]
+ \C^i(\k,\q)\; \tC^{l j}(\k,\q)
 \tr \left[\mathcal{U}(\y)\mathcal{U}^{l\;\dagger}_{[0,1]}(\y^\prime)\right]\nonumber\\
&&\hspace{-3.5cm} -i\; \C^j(\k,\q)\; \frac{\q^i}{\q^2}\, \tr\left[\mathcal{U}^\dagger(\y^\prime)\mathcal{U}_{[1,0]}(\y)\right]
+i\; \C^i(\k,\q)\; \frac{\q^j}{\q^2}\,
 \tr \left[\mathcal{U}(\y)\mathcal{U}^{\dagger}_{[1,0]}(\y^\prime)\right]
\Bigg]\, ,
\label{M2NE}
\eeq
where $\C^i(\k,\q)$ is defined in eq. \eqref{coefficient1} and $\tC^{i j}(\k,\q)$ is defined as
\beq
\tC^{i j}(\k,\q)&=&\left(\delta^{ij}-2\, \k^i\frac{\q^j}{\q^2}\right)\, .
\label{coefficient2}
\eeq
Finally, the next-to-next-to-eikonal contribution can be written as
\beq
&&\hspace{-1cm}\left. \overline{\mathcal{M}}^{ab}_{\lambda}(\uk,\q)^\dagger \overline{\mathcal{M}}^{ab}_{\lambda}(\uk,\q)\right|_{\rm NNE}=\varepsilon^{i*}_{\lambda}\varepsilon^{j}_{\lambda}\int d^2\y \int d^2\y^\prime \, e^{i(\y-\y^\prime)\cdot(\q-\k)} \;  2\left(\frac{L^+}{k^+}\right)^2 \nonumber\\
&&\hspace{-0.5cm}\times \Bigg\{   \C^j(\k,\q)
\Bigg[
-2\, \frac{\q^{i}}{\q^2}\, \k^l\k^m\, \tr\left[\mathcal{U}^\dagger(\y^\prime)\mathcal{U}^{lm}_{[0,2]}(\y)\right]
-i\, \frac{\q^{i}}{\q^2}\, \k^l\, \tr\left[\mathcal{U}^\dagger(\y^\prime)\mathcal{U}^{l}_{[1,1]}(\y)\right]
\nonumber \\
&& \hspace{1.9cm}
+\frac{1}{2}\, \frac{\q^{i}}{\q^2}\, \tr\left[\mathcal{U}^\dagger(\y^\prime)\mathcal{U}_{[2,0]}(\y)\right]
+\frac{i}{4}\, (\k^2\delta^{ij}-2\k^i\k^l)\, \tr\left[\mathcal{U}^\dagger(\y^\prime)\mathcal{U}^l_{(\rm A)}(\y)\right]
\nonumber\\
&&\hspace{1.9cm}
+\frac{\k^l}{4}\, \tr\left[\mathcal{U}^\dagger(\y^\prime)\mathcal{U}^{il}_{(\rm B)}(\y)\right]
+\frac{i}{4}\, \tr\left[\mathcal{U}^\dagger(\y^\prime)\mathcal{U}^{i}_{(\rm C)}(\y)\right] \Bigg]\nonumber \\
\nonumber \\
&&\hspace{-0.2cm}+\;
\C^i(\k,\q)
\Bigg[
-2\, \frac{\q^{j}}{\q^2}\, \k^l\k^m\,
\tr\left[\mathcal{U}(\y)\mathcal{U}^{lm\, \dagger}_{[0,2]}(\y^\prime)\right]
+i\, \frac{\q^{j}}{\q^2}\, \k^l\,
\tr\left[\mathcal{U}(\y)\mathcal{U}^{l\; \dagger}_{[1,1]}(\y^\prime)\right]
\nonumber \\
&&\hspace{1.9cm}
+\frac{1}{2}\, \frac{\q^{j}}{\q^2}\, \tr\left[\mathcal{U}(\y)\mathcal{U}^{\dagger}_{[2,0]}(\y^\prime)\right]
 -\frac{i}{4}\, (\k^2\delta^{jl}-2\k^{j}\k^l)\,
 \tr\left[\mathcal{U}(\y)\mathcal{U}^{l\, \dagger}_{(\rm A)}(\y^\prime)\right]
\nonumber \\
&&\hspace{1.9cm}
+\frac{\k^l}{4}\,
\tr\left[\mathcal{U}(\y)\mathcal{U}^{jl\; \dagger}_{(\rm B)}(\y^\prime)\right]
-\frac{i}{4}\,
\tr \left[\mathcal{U}(\y)\mathcal{U}^{j\; \dagger}_{(\rm C)}(\y^\prime)\right]
\Bigg]  \\
&&\hspace{-0.3cm} +\; \frac{1}{2}
\Bigg[
\tC^{l i}(\k,\q)\; \tC^{m j}(\k,\q)\;
\tr\left[\mathcal{U}^l_{[0,1]}(\y)\mathcal{U}^{m\; \dagger}_{[0,1]}(\y^\prime)\right]
+\frac{\q^{i}}{\q^2}\, \frac{\q^{j}}{\q^2}\, \tr\left[\mathcal{U}_{[1,0]}(\y)\mathcal{U}^{\dagger}_{[1,0]}(\y^\prime)\right] \nonumber\\
&&\hspace{0cm} -i\frac{\q^{i}}{\q^2}\, \tC^{l j}(\k,\q)
\tr\left[\mathcal{U}_{[1,0]}(\y)\mathcal{U}^{l\; \dagger}_{[0,1]}(\y^\prime)\right]
+i\frac{\q^{j}}{\q^2}\; \tC^{l i}(\k,\q)\;
\tr\left[\mathcal{U}^l_{[0,1]}(\y)\mathcal{U}^{\dagger}_{[1,0]}(\y^\prime)\right]
\Bigg] \Bigg\} . \nonumber
\label{M2NNE}
\eeq

The next step is to perform the averaging $\langle\cdots\rangle_A$ over the target fields.  Let us introduce the variables $\r=\y-\y'$ and $\b=\frac{1}{2}(\y+\y')$ and define the usual adjoint dipole
\beq
\O(\r)&=&\int d^2\b\;  \frac{1}{N_c^2-1}\left\langle\tr\left[\U\left(\b+\frac{\r}{2}\right)
\U^{\dagger}\left(\b-\frac{\r}{2}\right)\right]\right\rangle_A\nonumber\\
&=&\int d^2\b\;  \frac{1}{N_c^2-1}\left\langle\tr\left[\U^\dagger\left(\b+\frac{\r}{2}\right)\U\left(\b-\frac{\r}{2}\right)\right]\right\rangle_A
\label{adj_dip}\, ,
\eeq
the adjoint dipoles with one decorated Wilson line $\U_{[\alpha,\beta]}^{i\cdots j}$
\beq
\O_{[\alpha,\beta]}^{i\cdots j}(\r)&=&\int d^2\b\;  \frac{1}{N_c^2-1} \left\langle\tr\left[ \U_{[\alpha,\beta]}^{i\cdots j}\left(\b+\frac{\r}{2}\right)
\U^{\dagger}\left(\b-\frac{\r}{2}\right) \right]\right\rangle_A\nonumber\\
&=&\int d^2\b\;  \frac{1}{N_c^2-1} \left\langle\tr\left[
\U_{[\alpha,\beta]}^{i\cdots j\: \dagger}\left(\b+\frac{\r}{2}\right) \U\left(\b-\frac{\r}{2}\right)
 \right]\right\rangle_A\, ,
\label{adj_dec_dip}
\eeq
and analog operators for the case of the decorated Wilson lines $\U^{i}_{({\rm A})}$, $\U^{ij}_{({\rm B})}$ and $\U^{i}_{({\rm C})}$. Finally, let us also define the decorated adjoint dipoles formed from two decorated Wilson lines, as
\beq
\O_{[\alpha,\beta];[\gamma,\delta]}^{i\cdots j;l\cdots m}(\r)&=&\int d^2\b\;  \frac{1}{N_c^2-1} \left\langle\tr\left[
\U_{[\alpha,\beta]}^{i\cdots j}\left(\b+\frac{\r}{2}\right)
\U_{[\gamma,\delta]}^{l\cdots m\: \dagger}\left(\b-\frac{\r}{2}\right)  \right]\right\rangle_A\nonumber\\
&=&\int d^2\b\;  \frac{1}{N_c^2-1}\left\langle\tr\left[
\U_{[\alpha,\beta]}^{i\cdots j\: \dagger}\left(\b+\frac{\r}{2}\right)
\U_{[\gamma,\delta]}^{l\cdots m}\left(\b-\frac{\r}{2}\right)
\right]\right\rangle_A\, .
\label{adj_double_dec_dip}
\eeq
Here, $\langle\cdots\rangle_A$ stands for the averaging over the target fields. From the cyclicity of the trace in eq. \eqref{adj_double_dec_dip}, one obtains the identity
\beq
\O_{[\alpha,\beta];[\gamma,\delta]}^{i\cdots j;l\cdots m}(\r)
= \O_{[\gamma,\delta];[\alpha,\beta]}^{l\cdots m;i\cdots j}(-\r)
\label{transpose_identity_double_dec_dip}\, .
\eeq

Now, by using the definitions of the usual and the decorated dipole operators eqs. \eqref{adj_dip}, \eqref{adj_dec_dip} and \eqref{adj_double_dec_dip}, we can write the square of the gluon-nucleus reduced amplitude already averaged over the target fields separately as eikonal and beyond eikonal contributions. The eikonal contribution simply reads
\beq
\hspace{-1cm}\left. \frac{1}{N_c^2-1}\left\langle\overline{\mathcal{M}}^{ab}_{\lambda}(\uk,\q)^\dagger \overline{\mathcal{M}}^{ab}_{\lambda}(\uk,\q)\right\rangle\right |_{\rm \; E}\!\!\!=\varepsilon^{i*}_{\lambda}\varepsilon^{j}_{\lambda}\int d^2\r \; e^{i\r\cdot(\q-\k)}\, 4\, \C^i(\k,\q)\; \C^j(\k,\q)\; \O(\r)\, .
\eeq
Similarly, the next-to-eikonal contribution can be written as
\beq
\label{M2_avg_NE}
&&\hspace{-1cm}\left. \frac{1}{N_c^2-1}\left\langle \overline{\mathcal{M}}^{ab}_{\lambda}(\uk,\q)^\dagger \overline{\mathcal{M}}^{ab}_{\lambda}(\uk,\q)\right\rangle\right |_{\rm \; NE}\!\!\!=\varepsilon^{i*}_{\lambda}\varepsilon^{j}_{\lambda}\int d^2\r \; e^{i\r\cdot(\q-\k)} \, 2\, \frac{L^+}{k^+}\bigg\{ \C^j(\k,\q)\\
&&\hspace{-0.8cm}\times
\bigg[\tC^{l i}(\k,\q)\; \O^l_{[0,1]}(\r) -i\frac{\q^i}{\q^2}\, \O_{[1,0]}(\r)\bigg]
+\C^i(\k,\q)\, \bigg[\tC^{l j}(\k,\q)\; \O^l_{[0,1]}(-\r) +i\frac{\q^j}{\q^2}\, \O_{[1,0]}(-\r)\bigg]\bigg\}\, .\nonumber
\eeq
Finally, the next-to-next-to-eikonal contribution reads
\beq
&&\hspace{-1cm}\left. \frac{1}{N_c^2-1}\left\langle \overline{\mathcal{M}}^{ab}_{\lambda}(\uk,\q)^\dagger \overline{\mathcal{M}}^{ab}_{\lambda}(\uk,\q)\right\rangle\right |_{\rm \; NNE}\!\!\!=\varepsilon^{i*}_{\lambda}\varepsilon^{j}_{\lambda}\int d^2\r \; e^{i\r\cdot(\q-\k)} \, 2\,\left( \frac{L^+}{k^+}\right)^2 \nonumber\\
&& \hspace{1.2cm}\times\, \bigg\{\C^j(\k,\q) \bigg[
-2\, \frac{\q^{i}}{\q^2}\, \k^l\k^m\, \O^{lm}_{[0,2]}(\r) -i\, \frac{\q^{i}}{\q^2}\, \k^l\, \O^{l}_{[1,1]}(\r) +\frac{\q^i}{2\q^2}\, \O_{[2,0]}(\r)\nonumber\\
&&\hspace{3.7cm} +\frac{i}{4}\, \left(\k^2\delta^{il}-2\k^i\k^l\right)\,
\O^l_{(\rm A)}(\r) +\frac{\k^l}{4}\, \O^{il}_{(\rm B)}(\r)
+\frac{i}{4}\, \O^{i}_{(\rm C)}(\r)\bigg]\nonumber\\
&&\hspace{1.6cm}+\, \C^i(\k,\q)
\bigg[
-2\, \frac{\q^{j}}{\q^2}\, \k^l\k^m\, \O^{lm}_{[0,2]}(-\r)
+i\, \frac{\q^{j}}{\q^2}\, \k^l\, \O^{l}_{[1,1]}(-\r) +\frac{\q^j}{2\q^2}\, \O_{[2,0]}(-\r)\nonumber\\
&&\hspace{3.7cm} -\frac{i}{4}\, \left(\k^2\delta^{jl}-2\k^j\k^l\right)\,
 \O^l_{(\rm A)}(-\r) +\frac{\k^l}{4}\, \O^{jl}_{(\rm B)}(-\r)
 -\frac{i}{4}\, \O^{j}_{(\rm C)}(-\r)\bigg]\nonumber\\
&&\hspace{1.6cm} +\frac{1}{2}\, \bigg[\tC^{l i}(\k,\q)\; \tC^{m j}(\k,\q)\; \O^{l;m}_{[0,1];[0,1]}(\r) +\frac{\q^i}{\q^2}\, \frac{\q^j}{\q^2}\, \O_{[1,0];[1,0]}(\r)\nonumber\\
&&\hspace{2.3cm}-i\, \frac{\q^i}{\q^2}\, \tC^{l j}(\k,\q)\; \O^{l}_{[0,1];[1,0]}(-\r) +i\, \frac{\q^j}{\q^2}\, \tC^{l i}(\k,\q)\; \O^l_{[0,1];[1,0]}(\r)\bigg]\bigg\}
\, .\label{M2_avg_NNE}
\eeq

It is possible to further simplify eqs. \eqref{M2_avg_NE} and \eqref{M2_avg_NNE} by considering symmetry properties of the decorated dipole operators. The presence of the gluon background field $\A^{-}_a$ and of the projectile current $j^+_a$  explicitly breaks Lorentz invariance. But for the decorated dipole operators, which are independent of the projectile current, the averaging $\langle\cdots\rangle_A$ over the background field partially restores Lorentz symmetry. In particular, it restores the symmetry under rotations in the transverse plane, around the center of the target.

Hence, the decorated dipole operators should be covariant under such rotations. More precisely, operators like $\O_{[1,0]}(\r)$ or $\O_{[1,0];[1,0]}(\r)$, which have no transverse indices, should behave as scalars, whereas operators with one transverse index like $\O^i_{[0,1]}(\r)$ should behave as vectors, and the ones with more indices should behave as higher rank tensors. Moreover, since the $\b$ integration has already been taken in eqs. \eqref{adj_dip}, \eqref{adj_dec_dip} and \eqref{adj_double_dec_dip}, the decorated dipole operators depend only on one transverse vector, $\r$. Thus, under the transformation $\r \to -\r$ (which is a particular rotation in the transverse plane), the decorated dipole operators with an even number of transverse indices are invariant, whereas the ones with an odd number of indices flip sign.

By using these symmetry properties the next-to-eikonal, eq. \eqref{M2_avg_NE}, and next-to-next-to-eikonal, eq. \eqref{M2_avg_NNE}, contributions can be arranged as
\beq
\label{M2_avg_NE_Final-1}
&&\hspace{-0.3cm}\left. \frac{1}{N_c^2-1}\left\langle \overline{\mathcal{M}}^{ab}_{\lambda}(\uk,\q)^\dagger \overline{\mathcal{M}}^{ab}_{\lambda}(\uk,\q)\right\rangle\right |_{\rm \; NE}\!\!\!=\varepsilon^{i*}_{\lambda}\varepsilon^{j}_{\lambda}\int d^2\r \; e^{i\r\cdot(\q-\k)} \, 2\, \frac{L^+}{k^+}\\
&&\hspace{-0.3cm}\times
\bigg\{\!
\bigg[\C^{j}(\k,\q)\; \tC^{l i}(\k,\q) -\C^i(\k,\q)\; \tC^{l j}(\k,\q)\bigg]\, \O^{l}_{[0,1]}(\r)
 -i
\bigg[\C^j(\k,\q)\: \frac{\q^i}{\q^2}-\C^i(\k,\q)\: \frac{\q^j}{\q^2}\bigg]\, \O_{[1,0]}(\r) \! \bigg\}
\nonumber
\eeq
and
\beq
&&\hspace{-1cm}\left. \frac{1}{N_c^2-1}\left\langle \overline{\mathcal{M}}^{ab}_{\lambda}(\uk,\q)^\dagger \overline{\mathcal{M}}^{ab}_{\lambda}(\uk,\q)\right\rangle\right |_{\rm \; NNE}\!\!\!=\varepsilon^{i*}_{\lambda}\varepsilon^{j}_{\lambda}\int d^2\r \; e^{i\r\cdot(\q-\k)} \,\left( \frac{L^+}{k^+}\right)^2 \nonumber\\
&& \times\; \Bigg\{
\bigg[\C^j(\k,\q)\, \frac{\q^i}{\q^2} +\C^i(\k,\q)\, \frac{\q^j}{\q^2}\bigg] \bigg[-4\, \k^l\k^m\, \O^{l,m}_{[0,2]}(\r) -2i\, \k^l\, \O^{l}_{[1,1]}(\r) +\O_{[2,0]}(\r)\bigg]\nonumber\\
&&\hspace{0.7cm} +\frac{i}{2}\, \bigg[\C^j(\k,\q)\, (\k^2\delta^{il} -2\k^i\k^l) +\C^i(\k,\q)\, (\k^2\delta^{jl}-2\k^j\k^l)\bigg] \O^{l}_{(\rm A)}(\r)\nonumber\\
&&\hspace{0.7cm} +\frac{1}{2}\, \bigg[\C^j(\k,\q)\, \delta^{il} +\C^i(\k,\q)\, \delta^{jl}\bigg]\: \bigg[\k^m\, \O^{lm}_{(\rm B)}(\r)
+i\, \O^{l}_{(\rm C)}(\r)\bigg]\nonumber\\
&&\hspace{0.7cm} +\bigg[\tC^{l i}(\k,\q)\; \tC^{m j}(\k,\q)\; \O^{l;m}_{[0,1];[0,1]}(\r) +\frac{\q^i}{\q^2}\, \frac{\q^j}{\q^2}\, \O_{[1,0];[1,0]}(\r)\bigg]
\nonumber\\
&&\hspace{0.7cm} +i\bigg[\frac{\q^i}{\q^2}\, \tC^{l j}(\k,\q) +\frac{\q^j}{\q^2}\, \tC^{l i}(\k,\q)\bigg] \O^{l}_{[0,1];[1,0]}(\r)\Bigg\} \; .
\label{M2_avg_NNE_Final-2}
\eeq

The final expressions for the next-to-eikonal, eq. \eqref{M2_avg_NE_Final-1}, and next-to-next-to-eikonal, eq. \eqref{M2_avg_NNE_Final-2} contributions at the squared amplitude level are ready to be substituted to the aforementioned observables. Before we proceed further, we would like to point out an important realisation about the transverse momentum structures of the eikonal, next-to-eikonal and next-to-next-to eikonal terms.

The transverse momentum structure of the strict eikonal and next-to-next-to-eikonal terms are symmetric under the exchange of $i\leftrightarrow j$ unlike the next-to-eikonal terms whose transverse momentum structure is anti-symmetric under the same exchange. This affects directly the observables we are interested in.

In the calculation of the single inclusive gluon cross section one should sum over the gluon polarizations, which leads to
\beq
\sum_{\lambda}\varepsilon^{i*}_{\lambda}\varepsilon^j_{\lambda}=\delta^{ij}\, .
\eeq
Since the transverse momentum structure of the next-to-eikonal terms are antisymmetric under the exchange of $i\leftrightarrow j$, this contribution to the single inclusive gluon cross-section vanishes leading to the following result at next-to-next-to-eikonal accuracy
\beq
&&\hspace{-1.1cm}k^+ \frac{d\sigma}{dk^+\, d^2\k}= \int\frac{d^2\q}{(2\pi)^2} \; \varphi_p(\q) \, \frac{\q^2}{4}
\int d^2\r \; e^{i\r\cdot(\q-\k)}\; \Bigg\{4\, \C^i(\k,\q)\; \C^i(\k,\q)\, \O(\r)\nonumber\\
&&\hspace{-0.5cm}+\left( \frac{L^+}{k^+}\right)^2
\Bigg[2\, \frac{\q^i}{\q^2}\, \C^i(\k,\q)
\bigg[-4\, \k^l\k^m\, \O^{l,m}_{[0,2]}(\r) -2i\, \k^l\, \O^{l}_{[1,1]}(\r) +\O_{[2,0]}(\r)\bigg]\nonumber\\
&& 
+\C^i(\k,\q)
\bigg[ i(\k^2\delta^{il} -2\k^i\k^l)\, \O^{l}_{(\rm A)}(\r) +\k^m\, \O^{im}_{(\rm B)}(\r) +i\, \O^{i}_{(\rm C)}(\r)\bigg] \nonumber\\
&& 
+\tC^{l i}(\k,\q)
\bigg[\tC^{m i}(\k,\q)\; \O^{l;m}_{[0,1];[0,1]}(\r) +2i\, \frac{\q^i}{\q^2}\, \O^{l}_{[0,1];[1,0]}(\r)\bigg]
+\frac{1}{\q^2}\, \O_{[1,0];[1,0]}(\r)
\Bigg]\Bigg\}
\, .\label{Sigma_final}
\eeq

On the other hand, the light-front helicity asymmetry, as mentioned before, is calculated by taking the difference between the  $\lambda=+1$ and $\lambda=-1$ contributions. In this case, one uses
\beq
\sum_{\lambda} \; \lambda\; \varepsilon^{i*}_{\lambda}\; \varepsilon^j_{\lambda}=i\; \epsilon^{ij},
\eeq
where $\epsilon^{ij}$ is the antisymmetric matrix with $\epsilon^{12}=+1$.
Thus, the light-front helicity asymmetry gets the contribution from next-to-eikonal terms. The strict eikonal and next-to-next-to-eikonal terms vanish due to the fact that the transverse momentum structure for these terms are symmetric under the exchange of $i \leftrightarrow j$. The final expression of this asymmetry reads
\beq
&& \hspace{-0.3cm} k^+\, \frac{d\sigma^+}{dk^+\, d^2\k}-k^+\, \frac{d\sigma^-}{dk^+\, d^2\k}=\int \frac{d^2 \q}{(2\pi)^2}\;
 \varphi_p(\q)\; \q^2 \int d^2\r \; e^{i\r\cdot(\q-\k)} \,
 \nonumber\\
&&\hspace{0.3cm}\times
\left(\frac{L^+}{k^+}\right)  \epsilon^{ij}\, \C^j(\k,\q)  \bigg\{\!
 i\, \tC^{l i}(\k,\q)\, \O^{l}_{[0,1]}(\r)
 + \frac{\q^i}{\q^2}\, \O_{[1,0]}(\r) \! \bigg\} \; .
\label{M2_avg_NE_Final}
\eeq

\section{Conclusions}

The retarded gluon propagator in a classical  background field is one of the most important building blocks of the high energy dilute-dense scattering processes and also of medium-induced gluon radiation. Corrections beyond the eikonal approximation for this propagator were first considered in Ref. \cite{Altinoluk:2014oxa}. 
The corrections associated with the finite length of the medium were calculated at next-to-eikonal accuracy, meaning at order
$O\left(\frac{L^+}{k^+}\partial^2_{\perp}\right)$.
These corrections involve new operators referred to as decorated Wilson lines, as they include insertions of gradients of the background field along the path.

In this paper, we extend our study of finite width effects on the retarded gluon propagator, including next-to-next-to-eikonal corrections that are order $\left(\frac{L^+}{k^+}\partial^2_{\perp}\right)^2$. The new operators that appear at next-to-next-to-eikonal accuracy are also decorated Wilson lines but with higher number of insertions of the gradients of the background field or with higher derivatives.

The eikonal expansion performed  at the level of the gluon background propagator is then applied to high energy dilute-dense scattering processes within the CGC framework. Two different observables have been analysed, in pA collisions at midrapidity, within this framework: the single inclusive gluon production cross section and the light-front helicity asymmetry of  produced gluons. For the single inclusive gluon cross section, it has been shown that the next-to-eikonal terms vanish and the first non-vanishing corrections to the strict eikonal limit that appear at next-to-next-to-eikonal order have been calculated. On the other hand, for the light-front helicity asymmetry, it has been shown that both the strict eikonal terms and next-to-next-to-eikonal terms vanish and the leading contribution to this observable turns out to be the next-to-eikonal terms. This result shows the analogy between the twist expansion of the hard processes and the eikonal expansion of the high-energy processes.

The decorated dipole operators appearing in both of the observables at next-to-eikonal and next-to-next-eikonal accuracy are expected to have rapidity divergences. Therefore, understanding the low-$x$ evolution of these operators is a complementary extension of the analysis of the CGC beyond eikonal accuracy. This issue is left for future studies.

The eikonal expansion that is considered in detail in this paper has further applications. It can be applied to other high energy dilute-dense processes such as DIS and single inclusive gluon production in the hybrid formalism, or to jet quenching physics.
Such applications are also left for future studies.

\section*{Acknowledgements}
We thank Carlos Salgado for discussions on the subject, and
Eric Laenen for explanations of his works \cite{Laenen:2008gt,Laenen:2010uz}. This research  was supported by the People Programme (Marie Curie Actions) of the European Union's Seventh Framework Programme FP7/2007-2013/ under REA
grant agreement \#318921 (TA, NA and GB); the Kreitman Foundation and the ISRAELI SCIENCE FOUNDATION grant \#87277111 (GB);  the European Research Council grant HotLHC ERC-2011-StG-279579, Ministerio de Ciencia e Innovaci\'on of Spain under project FPA2011-22776, Xunta de Galicia (Conseller\'{\i}a de Educaci\'on and Conseller\'\i a de Innovaci\'on e Industria - Programa Incite),  the Spanish Consolider-Ingenio 2010 Programme CPAN and  FEDER (TA, NA and AM).

\appendix

\section{Details of the semi-classical expansion\label{sec:app_semi_class_exp}}

In this appendix, we calculate the semi-classical expansion of $\widetilde{\R}^{ab}_{\uk}(x^+,y^+;\y)$ up to order $1/{k^+}^2$ starting from the expression \eqref{finalRtilde}, using the method outlined in section \ref{sec:semi_class_exp}. For simplicity, we consider separately each term with a given index $l$ from the series representation \eqref{finalRtilde} of $\widetilde{\R}^{ab}_{\uk}(x^+,y^+;\y)$. Note that terms with $l>5$ can only give contributions more suppressed than $1/{k^+}^2$ in the $k^+\rightarrow +\infty$ limit.

\subsection{$l=1$ term in eq. \eqref{finalRtilde}}

The term with $l=1$ in eq. \eqref{finalRtilde} can be written as
\beq
&&\int_{y^+}^{x^+}dz^+\int d^2\u \; \G_{0,k^+}(z^+,\u;y^+,0)\nonumber\\
&&\times\P_+\Bigg\{ \U\left( x^+,y^+;[\z(z^+)]\right)ig\;T\cdot\left[\A^-(z^+,\hat{\z}(z^+)+\u)-\A^-(z^+,\hat{\z}(z^+))\right]\Bigg\}\nonumber\\
&&=\P_+\U\left( x^+,y^+;[\hat{\z}(z^+)]\right)\int_{y^+}^{x^+}dz^+\; ig\;T\cdot \int d^2\u \; \G_{0,k^+}(z^+,\u;y^+,0)\\
&&\times\bigg\{\bigg[\u^i\del_i+\frac{\u^i\u^j}{2!}\del_i\del_j+\frac{\u^i\u^j\u^l}{3!}\del_i\del_j\del_l+\frac{\u^i\u^j\u^l\u^m}{4!}\del_i\del_j\del_l\del_m\bigg] \A^-(z^+,\hat{\z}(z^+))\bigg\}\, .\nonumber
\eeq
Note that in this equation, each partial derivative is implicitly with respect to $\y$ and they are acting only on the background field $\A^-$. Also note that the free scalar propagator $\G_{0,k^+}$ is even in $\u$, so that any odd power of $\u$ appearing in the expansion vanishes. Moreover, due to azimuthal symmetry, we can make the replacements
\beq
\label{2u}
&&\u^i\u^j\mapsto \frac{\delta^{ij}}{2}\u^2\,,\\
\label{4u}
&&\u^i\u^j\u^l\u^m \mapsto \frac{(\u^2)^2}{8}(\delta^{ij}\delta^{lm}+\delta^{il}\delta^{jm}+\delta^{im}\delta^{jl})\, .
\eeq
Thus, the $l=1$ term in $\widetilde{\R}^{ab}_{\uk}(x^+,y^+;\y)$ can be written as
\beq
\label{}
&&\hspace{-2.5cm}\widetilde{\R}^{ab}_{\uk}(x^+,y^+;\y)_{l=1}= \P_+\; \U\left( x^+,y^+;[\hat{\z}(z^+)]\right)\,ig\, T\cdot \int_{y^+}^{x^+}dz^+\nonumber\\
&&\times\bigg\{ \; \frac{1}{4}\Big[\partial_{\perp}^2\A^-(z^+,\hat{\z}(z^+))\Big]
\int d^2\u\; (\u^2) \;\G_{0,k^+}(z^+,\u;y^+,0)
 \nonumber\\
&&
+ \frac{1}{64}\Big[(\partial_{\perp}^2)^2\A^-(z^+,\hat{\z}(z^+))\Big]\int d^2\u\; (\u^2)^2 \;\G_{0,k^+}(z^+,\u;y^+,0) \bigg\}.
\eeq
Using the expression \eqref{Free_scalar_propag} for the retarded free scalar propagator, it is straightforward to obtain the following relations :
\beq
\label{n1}
\int d^2\u  \;\G_{0,k^+}(z^+,\u;y^+,0) \; (\u^2) &=& 2i\;\frac{(z^+-y^+)}{k^+}\, ,\\
\label{n2}
\int d^2\u  \;\G_{0,k^+}(z^+,\u;y^+,0) \; (\u^2)^2 &=& -8\;\frac{(z^+-y^+)^2}{(k^+)^2}\, .
\eeq
Then, by using eqs. \eqref{n1} and \eqref{n2}, the $l=1$ contribution to $\widetilde{\R}^{ab}_{\uk}(x^+,y^+;\y)$ reads
\beq
\label{R0}
\widetilde{\R}^{ab}_{\uk}(x^+,y^+;\y)_{l=1}&=&  \P_+\; \U\left( x^+,y^+;[\hat{\z}(z^+)]\right)\,ig\, T\cdot \int_{y^+}^{x^+}dz^+\bigg\{ i\frac{(z^+-y^+)}{2k^+}\Big[\partial_{\perp}^2\A^-(z^+,\hat{\z}(z^+))\Big]\nonumber\\
&-&\frac{(z^+-y^+)^2}{8(k^+)^2}\Big[(\partial_{\perp}^2)^2\A^-(z^+,\hat{\z}(z^+))\Big]\bigg\}
\eeq
or, using the notations introduced in Eqs. \eqref{Bij} and \eqref{Bijlm},
\beq
\label{Rl1final}
&&
\hspace{-1cm}
\widetilde{\R}^{ab}_{\uk}(x^+,y^+;\y)_{l=1}=  \P_+\; \U\left( x^+,y^+;[\hat{\z}(z^+)]\right)
\nonumber\\
&&\hspace{-0.5cm}
\times
\int_{y^+}^{x^+}dz_1^+\bigg\{ i\frac{(z_1^+\!-\!y^+)}{2k^+}\delta^{ij}\, {\cal B}^{ij}(\mz_1)
-\frac{(z_1^+\!-\!y^+)^2}{8(k^+)^2}\,
\delta^{ij}\delta^{lm}\: {\cal B}^{ijlm}(\mz_1)
\bigg\}.
\eeq

\subsection{$l=2$ term in eq. \eqref{finalRtilde}}

The term with $l=2$ in eq. \eqref{finalRtilde} reads
\beq
\label{Rl2first}
&&\hspace{-1cm}\widetilde{\R}^{ab}_{\uk}(x^+,y^+;\y)_{l=2}=\int_{y^+}^{x^+}dz^+_1\int_{z^+_1}^{x^+}dz^+_2\int d^2\u_1\int d^2\u_2 \;\G_{0,k^+}(z^+_2,\u_2;z^+_1,\u_1)  \nonumber\\
&&
\times\;
 \P_+\; \U\left( x^+,y^+;[\hat{\z}(z^+)]\right)\Big[igT\cdot\delta\A^-_2(\u_2)\Big]\Big[igT\cdot\delta\A^-_1(\u_1)\Big]
\G_{0,k^+}(z^+_1,\u_1;y^+,0)\nonumber\\
&&\hspace{2cm}=\int_{y^+}^{x^+}dz^+_1\int_{z^+_1}^{x^+}dz^+_2\int d^2\u_1\int d^2\r \;\G_{0,k^+}(z^+_2,\r;z^+_1,0)  \\
&&
\hspace{-0.5cm}\times\;
 \P_+\; \U\left( x^+,y^+;[\hat{\z}(z^+)]\right)\Big[igT\cdot\delta\A^-_2(\u_1+\r)\Big]\Big[igT\cdot\delta\A^-_1(\u_1)\Big]
\G_{0,k^+}(z^+_1,\u_1;y^+,0)\, , \nonumber
\eeq
where $\r\equiv \u_2-\u_1$. Taylor expanding the $\delta\A^-$s with respect to $\u_1$ and $\r$, and dropping the terms vanishing due to azimuthal symmetry, eq. \eqref{Rl2first} becomes
\beq
\label{Rl2second}
&&\widetilde{\R}^{ab}_{\uk}(x^+,y^+;\y)_{l=2}\nonumber\\ &=&\P_+\; \U\left( x^+,y^+;[\hat{\z}(z^+)]\right)\int_{y^+}^{x^+}\!\!\!\!\!dz^+_1\int_{z^+_1}^{x^+}\!\!\!\!\!dz^+_2\int \!\!d^2\u_1\int \!\!d^2\r \;\G_{0,k^+}(z^+_2,\r;z^+_1,0) \;\nonumber\\
&\times& \G_{0,k^+}(z^+_1,\u_1;y^+,0)
\bigg\{
\u_1^i\; {\cal B}^i(\mz_2)\; \u_1^j\; {\cal B}^j(\mz_1)
+\bigg[\frac{1}{6}\u_1^i\u_1^j\u_1^l+\frac{1}{2}\r^i\r^j\u_1^l\bigg]\;{\cal B}^{ijl}(\mz_2)\;
\u^m_1\; {\cal B}^m(\mz_1)
\nonumber\\
&
+&\frac{1}{2}\bigg[\u_1^i\u_1^j+\r^i\r^j\bigg]\; {\cal B}^{ij}(\mz_2)\;
\frac{1}{2}\u_1^l\u_1^m\; {\cal B}^{lm}(\mz_1)
+\u_1^i\; {\cal B}^i(\mz_2)\;
\frac{1}{6}\u_1^j\u_1^l\u_1^m\; {\cal B}^{jlm}(\mz_1)
\bigg\}\, .
\eeq
Then, using eqs. \eqref{2u} and \eqref{4u} and integrating over $\u_1$ and $\r$ by  eqs. \eqref{n0},  \eqref{n1} and  \eqref{n2}, we get
\beq
\label{Rl2final}
\widetilde{\R}^{ab}_{\uk}(x^+,y^+;\y)_{l=2} &=& \P_+\; \U\left( x^+,y^+;[\hat{\z}(z^+)]\right)\,\int_{y^+}^{x^+}dz^+_1\int_{z^+_1}^{x^+}dz^+_2
\bigg\{ i\frac{(z^+_1\!-\!y^+)}{k^+}\, {\cal B}^i(\mz_2)\, {\cal B}^i(\mz_1)
\nonumber\\
&&
-\frac{(z^+_2\!-\!y^+)(z^+_1\!-\!y^+)}{2(k^+)^2}
\Big[\delta^{jl}\, {\cal B}^{ijl}(\mz_2)\: {\cal B}^i(\mz_1)
+\frac{1}{2}\delta^{ij}\, {\cal B}^{ij}(\mz_2)\: \delta^{lm}\, {\cal B}^{lm}(\mz_1)\Big]
\nonumber\\
&&-\frac{(z^+_1\!-\!y^+)^2}{2(k^+)^2}\,\Big[
{\cal B}^{i}(\mz_2)\: \delta^{jl}\, {\cal B}^{ijl}(\mz_1)
+{\cal B}^{ij}(\mz_2)\: {\cal B}^{ij}(\mz_1)\Big]
\bigg\}\, .
\eeq


\subsection{$l=3$ term in eq. \eqref{finalRtilde}}

For $l=3$, one has
\beq
\label{Rl3first}
&&\widetilde{\R}^{ab}_{\uk}(x^+,y^+;\y)_{l=3}=\P_+\; \U \left( x^+,y^+;[\hat{\z}(z^+)]\right)\,
\int_{y^+}^{x^+}dz^+_1 \int_{z^+_1}^{x^+}dz^+_2 \int_{z^+_2}^{x^+}dz^+_3
\nonumber\\
&&
\times \,
\int d^2\u_1\int d^2\u_2 \int d^2\u_3\;
 \G_{0,k^+}(z^+_3,\u_3;z^+_2,\u_2)\; \G_{0,k^+}(z^+_2,\u_2;z^+_1,\u_1)\;
 \nonumber\\
 &&\times \;
  \G_{0,k^+}(z^+_1,\u_1;y^+,0)\;\;
igT\cdot\delta\A^-_{3}(\u_3) \;\; igT\cdot\delta\A^-_{2}(\u_2) \;\; igT\cdot\delta\A^-_{1}(\u_1)
\nonumber\\
&&\hspace{3cm}=\P_+\; \U \left( x^+,y^+;[\hat{\z}(z^+)]\right)\,
\int_{y^+}^{x^+}dz^+_1 \int_{z^+_1}^{x^+}dz^+_2 \int_{z^+_2}^{x^+}dz^+_3
\nonumber\\
&& \times
\int d^2\s \int d^2\r \int d^2\u\;
\G_{0,k^+}(z^+_3,\s;z^+_2,0)\; \G_{0,k^+}(z^+_2,\r;z^+_1,0)\; \G_{0,k^+}(z^+_1,\u;y^+,0)\,
\nonumber\\
&&\times
\;\; igT\cdot\delta\A^-_{3}(\s+\r+\u) \;\; igT\cdot\delta\A^-_{2}(\r+\u) \;\; igT\cdot\delta\A^-_{1}(\u)\, ,
\eeq
where $\s\equiv \u_3-\u_2$, $\r\equiv \u_2-\u_1$ and $\u\equiv \u_1$.
Taylor expanding with respect to $\s$, $\r$ and $\u$, and dropping the terms integrating to zero, one gets
\beq
\label{Rl3second}
&&\widetilde{\R}^{ab}_{\uk}(x^+,y^+;\y)_{l=3}=\P_+\; \U \left( x^+,y^+;[\hat{\z}(z^+)]\right)\,
\int_{y^+}^{x^+}dz^+_1\int_{z^+_1}^{x^+}dz^+_2\int_{z^+_2}^{x^+}dz^+_3
\nonumber\\
&& \times \;
\int d^2\s \int d^2\r \int d^2\u\;
 \G_{0,k^+}(z^+_3,\s;z^+_2,0)\; \G_{0,k^+}(z^+_2,\r;z^+_1,0)\; \G_{0,k^+}(z^+_1,\u;y^+,0)
 \nonumber\\
 &&\times \;
\bigg\{
\frac{1}{2}\u^i\u^j\u^l\u^m
\Big[
{\cal B}^{ij}(\mz_3)\; {\cal B}^l(\mz_2)\; {\cal B}^m(\mz_1)
+{\cal B}^i(\mz_3)\; {\cal B}^{jl}(\mz_2)\; {\cal B}^m(\mz_1)
+{\cal B}^i(\mz_3)\; {\cal B}^j(\mz_2)\; {\cal B}^{lm}(\mz_1)
\Big]
\nonumber\\
&&\hspace{0.5cm}
+\r^i\r^j\u^l\u^m
\Big[
{\cal B}^{il}(\mz_3)\; {\cal B}^j(\mz_2) +\frac{1}{2}{\cal B}^{ij}(\mz_3)\; {\cal B}^l(\mz_2)
+\frac{1}{2}{\cal B}^l(\mz_3)\; {\cal B}^{ij}(\mz_2)+{\cal B}^{i}(\mz_3)\; {\cal B}^{jl}(\mz_2)
\Big]\; {\cal B}^m(\mz_1)
\nonumber\\
&&\hspace{0.5cm}
+ \frac{1}{2}\s^i\s^j\u^l\u^m\; {\cal B}^{ij}(\mz_3)\; {\cal B}^l(\mz_2)\; {\cal B}^m(\mz_1)
+\frac{1}{2}\r^i\r^j\u^l\u^m\; {\cal B}^{i}(\mz_3)\; {\cal B}^{j}(\mz_2)\; {\cal B}^{lm}(\mz_1)
\bigg\}\, .
\eeq

Using eqs.  \eqref{2u} and \eqref{4u} and integrating over $\u$, $\r$ and $\s$ by eqs. \eqref{n0},  \eqref{n1} and  \eqref{n2}, one gets
\beq
\label{Rl3final}
&&\widetilde{\R}^{ab}_{\uk}(x^+,y^+;\y)_{l=3}=\P_+\; \U \left( x^+,y^+;[\hat{\z}(z^+)]\right)\,
\int_{y^+}^{x^+}dz^+_1\int_{z^+_1}^{x^+}dz^+_2\int_{z^+_2}^{x^+}dz^+_3
\nonumber\\
&&\times
\bigg\{ -\frac{(z^+_3\!-\!y^+)(z^+_1\!-\!y^+)}{2(k^+)^2}\delta^{ij}{\cal B}^{ij}(\mz_3)\; {\cal B}^l(\mz_2)\; {\cal B}^l(\mz_1)
-\frac{(z^+_1\!-\!y^+)^2}{(k^+)^2}{\cal B}^{i}(\mz_3)\; {\cal B}^{j}(\mz_2)\; {\cal B}^{ij}(\mz_1)
\nonumber\\
&&\hspace{0.7cm}
-\frac{(z^+_2\!-\!y^+)(z^+_1\!-\!y^+)}{2(k^+)^2}
\Big[
{\cal B}^{i}(\mz_3)\; \delta^{jl}{\cal B}^{jl}(\mz_2)\; {\cal B}^{i}(\mz_1)
+{\cal B}^{i}(\mz_3)\; {\cal B}^{i}(\mz_2)\; \delta^{jl}{\cal B}^{jl}(\mz_1)
\nonumber\\
&&\hspace{4.2cm}
+2\; {\cal B}^{ij}(\mz_3)\; {\cal B}^{i}(\mz_2)\; {\cal B}^j(\mz_1)
+2\; {\cal B}^{i}(\mz_3)\; {\cal B}^{ij}(\mz_2)\; {\cal B}^{j}(\mz_1)
\Big]\bigg\}\, .
\eeq


\subsection{$l=4$ term in eq. \eqref{finalRtilde}}

Finally, for $l=4$, one has
\beq
\label{Rl4first}
&&
\hspace{-0.8cm}
\widetilde{\R}^{ab}_{\uk}(x^+,y^+;\y)_{l=4}=\P_+\; \U \left( x^+,y^+;[\hat{\z}(z^+)]\right)\, \int_{y^+}^{x^+}dz^+_1 \int_{z^+_1}^{x^+}dz^+_2
\nonumber\\
&&\times\;
\int_{z^+_2}^{x^+}dz^+_3 \int_{z^+_3}^{x^+}dz^+_4
\int d^2\u_1 \int d^2\u_2 \int d^2\u_3 \int d^2\u_4 \; \G_{0,k^+}(z^+_4,\u_4;z^+_3,\u_3)
\nonumber\\
&&\times\;
\G_{0,k^+}(z^+_3,\u_3;z^+_2,\u_2)\;\;\G_{0,k^+}(z^+_2,\u_2;z^+_1,\u_1)
\;\;
 \G_{0,k^+}(z^+_1,\u_1;y^+,0)
 \nonumber\\
 &&\times\;
igT\cdot\delta\A^-_{4}(\u_4) \;\; igT\cdot\delta\A^-_{3}(\u_3) \;\; igT\cdot\delta\A^-_{2}(\u_2)\;\;
 igT\cdot\delta\A^-_{1}(\u_1)
 \nonumber\\
 &&
\hspace{-1cm}
\hspace{3cm}=\P_+\; \U \left( x^+,y^+;[\hat{\z}(z^+)]\right)\, \int_{y^+}^{x^+}dz^+_1  \int_{z^+_1}^{x^+}dz^+_2
\int_{z^+_2}^{x^+}dz^+_3  \int_{z^+_3}^{x^+}dz^+_4
\nonumber\\
&&
\times
\int d^2\t \int d^2\s \int d^2\r \int d^2\u \;\; \G_{0,k^+}(z^+_4,\t;z^+_3,0)\;\;
\G_{0,k^+}(z^+_3,\s;z^+_2,0)
\nonumber\\
&&\times
\;\; \G_{0,k^+}(z^+_2,\r;z^+_1,0)\;\; \G_{0,k^+}(z^+_1,\u;y^+,0)\;\; igT\cdot\delta\A^-_{4}(\t+\s+\r+\u)\nonumber\\
&&\times   \;\; igT\cdot\delta\A^-_{3}(\s+\r+\u) \;\; igT\cdot\delta\A^-_{2}(\r+\u)\;\;  igT\cdot\delta\A^-_{1}(\u)\, .
\eeq
Following the same steps as in the previous sections (and introducing $\t\equiv \u_4-\u_3$), one obtains
\beq
\label{Rl4second}
&&
\hspace{-1cm}
\widetilde{\R}^{ab}_{\uk}(x^+,y^+;\y)_{l=4}=\P_+\; \U \left( x^+,y^+;[\hat{\z}(z^+)]\right)\,
\int_{y^+}^{x^+}dz^+_1 \int_{z^+_1}^{x^+}dz^+_2 \int_{z^+_2}^{x^+}dz^+_3 \int_{z^+_3}^{x^+}dz^+_4
\nonumber\\
&&\times
\int d^2\t \int d^2\s \int d^2\r \int d^2\u\;\; \G_{0,k^+}(z^+_4,\t;z^+_3,0)\;\; \G_{0,k^+}(z^+_3,\s;z^+_2,0)\;
\nonumber\\
&&\times
\;\;  \G_{0,k^+}(z^+_2,\r;z^+_1,0)\;\; \G_{0,k^+}(z^+_1,\u;y^+,0)\;\; {\cal B}^{i}(\mz_4)\;\; {\cal B}^j(\mz_3)\;\; {\cal B}^l(\mz_2)\;\; {\cal B}^m(\mz_1)
\nonumber\\
&&\times\;
\Big[ \s^i\s^j\u^l\u^m + \u^i\u^j\u^l\u^m + \r^i\r^j\u^l\u^m + \r^i\u^j\r^l\u^m + \u^i\r^j\r^l\u^m\Big]\, ,
\eeq
and, finally,
\beq
\label{Rl4final}
&&
\hspace{-2cm}
\widetilde{\R}^{ab}_{\uk}(x^+,y^+;\y)_{l=4}=\P_+\; \U \left( x^+,y^+;[\hat{\z}(z^+)]\right)\,
\int_{y^+}^{x^+}dz^+_1 \int_{z^+_1}^{x^+}dz^+_2 \int_{z^+_2}^{x^+}dz^+_3 \int_{z^+_3}^{x^+}dz^+_4\nonumber\\
&&\times \; \bigg\{ -\frac{(z^+_3\!-\!y^+)(z^+_1\!-\!y^+)}{(k^+)^2}\;
{\cal B}^{i}(\mz_4)\;\; {\cal B}^i(\mz_3)\;\; {\cal B}^{j}(\mz_2)\;\; {\cal B}^{j}(\mz_1)
\nonumber\\
&&\hspace{0.8cm}
-\frac{(z^+_2\!-\!y^+)(z^+_1\!-\!y^+)}{(k^+)^2}\;
 {\cal B}^{i}(\mz_4)\;\; {\cal B}^j(\mz_3)\;\; {\cal B}^i(\mz_2)\;\; {\cal B}^j(\mz_1)
\nonumber\\
&&\hspace{0.8cm}
-\frac{(z^+_2\!-\!y^+)(z^+_1\!-\!y^+)}{(k^+)^2}\;
 {\cal B}^i(\mz_4)\;\; {\cal B}^j(\mz_3)\;\; {\cal B}^j(\mz_2)\;\; {\cal B}^i(\mz_1)
\bigg\}\, .
\eeq

\bibliography{biblio}{}
\bibliographystyle{jhep}

\end{document}